\documentclass{comjnl}
\usepackage{cite}
\usepackage{amsmath}
\usepackage{mathtools}
\usepackage{amssymb}
\usepackage{algorithm}
\usepackage{subcaption}
\usepackage[noend]{algpseudocode}
\usepackage[utf8]{inputenc}
\usepackage{float}

\algblock{ParFor}{EndParFor}
\algnewcommand\algorithmicparfor{\textbf{parfor}}
\algnewcommand\algorithmicpardo{\textbf{do}}
\algnewcommand\algorithmicendparfor{\textbf{end\ parfor}}
\algrenewtext{ParFor}[1]{\algorithmicparfor\ #1\ \algorithmicpardo}
\algrenewtext{EndParFor}{\algorithmicendparfor}

\algnewcommand\algorithmicforeach{\textbf{for each}}
\algdef{S}[FOR]{ForEach}[1]{\algorithmicforeach\ #1\ \algorithmicdo}

\usepackage{enumitem}

\usepackage{url}

\usepackage{booktabs}
\usepackage[table, xcdraw]{xcolor}

\begin{document}
\title{A Proof of Useful Work for Artificial Intelligence on the Blockchain}

\author{Andrei Lihu}
\affiliation{Up and Running Software} \email{andrei.lihu@upandrunningsoftware.com, jincheng@oben.com, igor.barjaktarevic@upandrunningsoftware.com,\\ patrick.gerzanics@upandrunningsoftware.com, mark@oben.com}
\author{Jincheng Du}
\affiliation{Oben}

\author{Igor Barjaktarevi\'{c}}
\affiliation{Up and Running Software}

\author{\\Patrick Gerzanics}
\affiliation{Up and Running Software}

\author{Mark Harvilla}
\affiliation{Oben}

\shortauthors{A. Lihu, J. Du, I. Barjaktarevi\'{c}, P. Gerzanics, M. Harvilla}

\begin{abstract}
Bitcoin mining is a wasteful and resource-intensive process. To add a block of transactions to the blockchain, miners spend a considerable amount of energy. The Bitcoin protocol, named ‘proof of work’ (PoW), resembles a lottery and the underlying computational work is not useful otherwise. In this paper, we describe a novel ‘proof of useful work’ (PoUW) protocol based on training a machine learning model on the blockchain. Miners get a chance to create new coins after performing honest ML training work. Clients submit tasks and pay all training contributors. This is an extra incentive to participate in the network because the system does not rely only on the lottery procedure. Using our consensus protocol, interested parties can order, complete, and verify useful work in a distributed environment. We outline mechanisms to reward useful work and punish malicious actors. We aim to build better AI systems using the security of the blockchain.
\end{abstract}

\maketitle

\section{Introduction}\label{sec:introduction}

Bitcoin \cite{nakamoto2012bitcoin} miners are spending a huge amount of electricity to append a new block to the blockchain. A blockchain is a sequence of blocks, where each block packs several transactions. Among other fields, the header of a block contains a cryptographic hash of the previous block, a target value, and a nonce. To add a new block, in the classical \emph{proof of work} (PoW) protocol, the \emph{double SHA-256 hash} of the block's header must fall under a target value which is known to all network participants. The nonce is varied to obtain different hash values. For each successfully mined block, a miner is allowed to create out of thin air a specific amount of new coins (the block subsidy) and also to collect the fees for transactions included in the block. Sometimes new blocks are created on top of the same previous block, which leads to forks. According to the consensus rule, miners should build on the longest chain, therefore shorter forks will be discarded. PoW is easy to verify, but hard to produce \cite{Ball2017}. Repeated hashing (by varying the nonce) is not useful \emph{per se}, PoW is a pure lottery and does not guarantee fair payment for all actors that spend computational resources. We would like to reward only productive actors. To do so, PoW should be replaced or combined with beneficial work (e.g. \emph{machine learning}) and miners should compete to provide a \emph{proof of useful work} (PoUW).

There are several obstacles in designing a PoUW system using machine learning (ML) as the basis for useful work. Compared to hashing, ML tasks are complex and diverse. The Bitcoin puzzle \cite{Ball2017} grows in complexity over time, but in a real-world scenario, the client's ML problem dictates the complexity. Due to the ML task's heterogeneity, it is difficult to verify if an actor performed honest work. There are many steps in the ML algorithms where bad actors can avoid work, perform cheap work, or behave maliciously. The blockchain is not designed to hold the large amounts of data needed for ML training and most of data is not of interest to the majority of actors. It is also hard to distribute and coordinate a ML training process in a trust-less environment as the blockchain's peer-to-peer (P2P) network.

A significant number of commercial machine learning platforms distribute and execute operations in parallel (e.g. Facebook \cite{DBLP:journals/corr/GoyalDGNWKTJH17}, Amazon \cite{Strom2015}, Google \cite{Dean:2012:LSD:2999134.2999271}). They work with Big Data that requires distribution of workloads and orchestration across multiple nodes. These systems use data parallelism: each node holds an internal replica of the trained ML model, while data is divided across worker nodes \cite{skymind_2018}. The blockchain offers access to more computational resources than a standard cloud ML service.

Our research was motivated by the questions: \emph{Can we build better AI systems using blockchain's security? Can we provide a better blockchain protocol based on PoUW?}

We used blockchain's infrastructure to coordinate the training of a deep neural network (DNN). We designed a decentralised network with a consensus protocol to perform and verify useful work based on ML. At the core of the protocol lies a novel way to create nonces derived from useful work.

Our paper is organised as follows: In Section \ref{Related work} we briefly present related efforts to combine artificial intelligence with the blockchain. We provide an overview of our environment and the protocol in Section \ref{Overview}. We show the detailed steps of the ML training in Section \ref{ML workflow}. In the section about the proof of useful work (Section \ref{PoUW}), we describe the mining and the verification processes. We also implemented a proof of concept (PoC) whose details are given in Section \ref{Implementation}. In Section \ref{Discussion}, we discuss potential performance and security concerns related to our proposal. We provide conclusions and details about future work in Section \ref{Conclusion}.

\section{Related work}\label{Related work}
A series of different PoUW protocols have been proposed, all of them with limited practicality: Ball et al. \cite{Ball2017} proposed methods for solving the \emph{orthogonal vectors}, \emph{all-pairs shortest path} and \emph{3SUM} problems on blockchain; in PrimeCoin miners search for Cunningham chains of prime numbers \cite{PrimeCoin}.

DML \cite{dml} and SingularityNET \cite{singularity} are marketplaces for distributed AI services utilizing smart contracts. SingularityNET requires curation of services. None of them have a tight integration with the blockchain, keeping AI as a separate service.

Gridcoin \cite{gridcoin} is an open-source proof-of-stake (PoS) blockchain for solving tasks on Berkeley Open Infrastructure for Network Computing and their system is centralised (whitelists, central server, centralised verification).

Coin.AI \cite{Baldominos} is just a theoretical proposal where miners train separately a DNN. The architecture of the deep neural network is generated based on the last mined block and the verification involves checking the model's performance. Their system is prone to security risks, e.g. miners doing cheap work or no work.

CrowdBC \cite{CrowdBC} is a crowdsourcing framework using Ethereum smart contracts. Participants deposit a certain monetary amount to mitigate for potential bad behaviour, but miners do not perform useful work.

According to our knowledge, our PoUW proposal is the only protocol that tightly integrates with the blockchain. It does not require smart contracts. The miners perform useful computational work that can be deterministically verified. Our PoUW facilitates better AI through decentralisation, enhanced security and the right incentives.

\section{System Overview} \label{Overview}
\subsection{Environment}
The environment is the PAI (Personalised Artificial Intelligence) blockchain, a hybrid Proof of Work/Proof of Stake (PoW/PoS) blockchain. It is a P2P decentralised network (\ref{fig:overview}) composed of:

\begin{figure}[!t]
    \centering
    \includegraphics[width=0.48\textwidth]{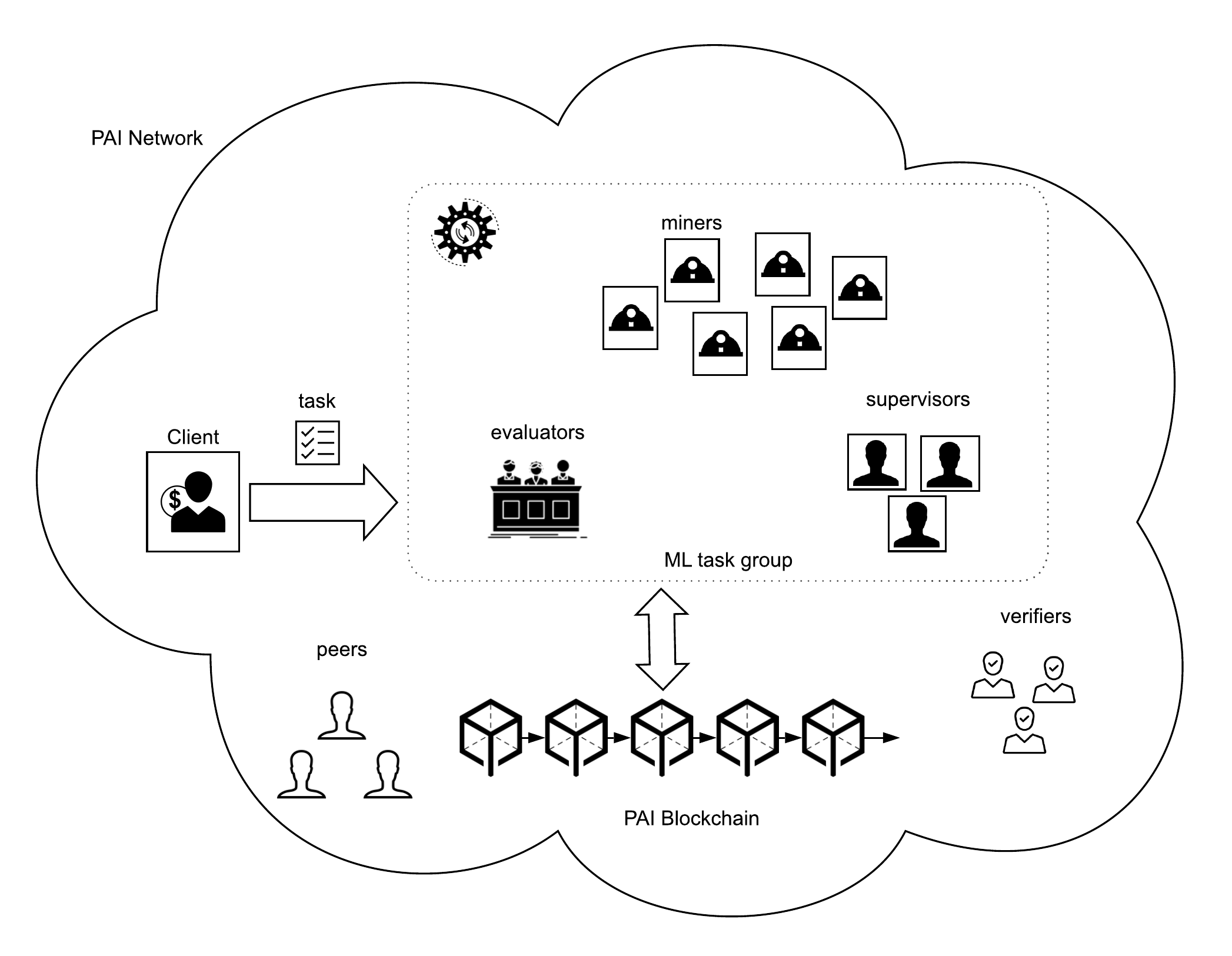}
    \caption{\textbf{Environment overview.} The client submits a ML task to the PAI network. Worker nodes perform the training and evaluators decide how to pay. The PAI blockchain ensures the security of the ML process.}
    \label{fig:overview}
\end{figure}

\begin{description}
	\item [Clients:] Nodes that pay to train their models on the PAI blockchain.
	\item [Miners:] Nodes that perform the training. They can mine a new block with special nonces obtained after each iteration. The training is distributed and all miners collaborate by sharing updates about their local model.
	\item [Supervisors:] Actors that record all messages during a task in a log called 'message history'. They also guard against malicious behaviour during the training because the environment may also contain Byzantine nodes.
	\item [Evaluators:] Independent nodes that test the final models from each miner and send the best one to the client. They also split the client's fee and pay all nodes accordingly.
	\item [Verifiers:] Nodes that verify if a block is valid. We need them because verification is computationally expensive and it is not carried out by all nodes.
	\item [Peers:] Nodes that do not have any of the aforementioned roles. They are using regular blockchain transactions.
\end{description}

\begin{figure}[!t]
    \centering
    \includegraphics[width=0.48\textwidth]{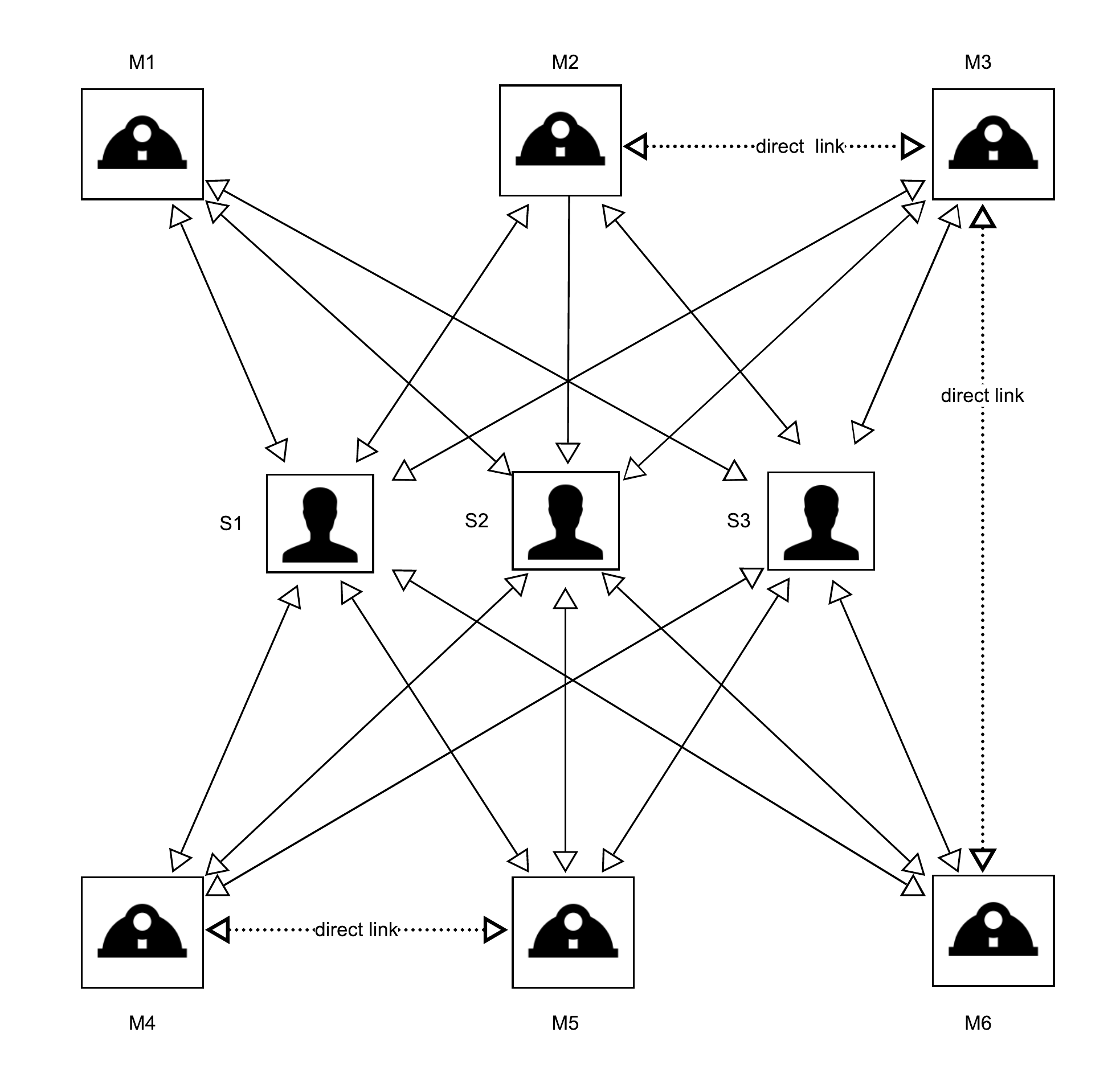}
    \caption{\textbf{A communication topology example.} Miners (M1-M6) are fully connected with the supervisors (S1-S3) and send/query data to/from them. A few miners established direct links in-between them (e.g. M2-M3, M4-M5, M3-M6).}
    \label{fig:communication}
\end{figure}

Miners and supervisors are called \emph{worker nodes} because they actively participate in the training. Worker nodes communicate by using fast message channels. To facilitate direct communication, we recommend running all worker nodes in a \emph{virtual private network} (VPN) with a full VPN mesh topology (e.g. PeerVPN -- \url{https://peervpn.net}). Miners send messages to supervisors that record the message history. Directly sending messages between miners is optional (see Fig. \ref{fig:communication}).

\subsection{Transactions}
A Bitcoin transaction is a digitally signed data structure that contains versioning, inputs and outputs. A transaction is valid if its inputs refer to other unspent outputs (UTXOs). \textbf{OP\_RETURN} is a script operation field (80 bytes in length in Bitcoin) that marks a transaction output invalid, but can be used to store arbitrary data in the blockchain.

In the PAI blockchain, we use special transactions to handle the training, verification, evaluation and payment of the ML tasks. To do so, we encode all the required extra information in the OP\_RETURN code (160 bytes in our protocol) of one of the transaction's outputs \footnote{We provided an example implementation for special transactions in the Supplementary material in \ref{appendix_transactions}, and in \ref{shortening} we detailed a method to shorten the task wait time.}.

Before inclusion in blocks, transactions wait in the \emph{mempool}, a buffer of pending transactions. Nodes relay transactions from the mempool to each other and eventually they propagate across the whole P2P network.

Off-chain messages and own transactions are signed using the participant's private ECDSA key. When we need multi-party transactions we use \emph{Boneh-Lynn-Shacham} (BLS) keys and signatures (\cite{Boneh2004}), for example when supervisors must agree on a common decision. To reduce the number of special transactions, we require that a \emph{leader} posts a multi-party transaction.

\subsection{Staking}
Except regular peers, all nodes must first deposit coins as a collateral. We call this process '\emph{staking}'. The deposits represent locked money that are later returned along with extra fees if the participants finish properly their work.

Staking means buying \emph{tickets}. Nodes can issue \emph{BUY\_TICKETS} transactions, which are special transactions containing the desired role (miner, supervisor etc) and preferences for specific tasks (encoded in the OP\_RETURN field). Tickets become 'live' after they are included in the blockchain and buried under a preset number of blocks (e.g. 128). There are around 40960 tickets in the \emph{mempool} and their number is kept constant by updating ticket prices after every 144 mined blocks. We use Decred's algorithm (\cite{Decred}) to adjust prices. No more than 50 tickets can be included in a mined block.

Another type of stake transaction is \emph{PAY\_FOR\_TASK}, which is issued by clients when they submit a new task. It includes a task description and a training fee.

Staking is a way to ensure against malicious behaviour because the stakes of dishonest actors are captured and re-distributed to fair players. At the end of training, the evaluators decide how to split the client's fee and punish the bad actors. Evaluators issue a \emph{CHARGE\_FOR\_TASK} transaction to pay the honest participants. The client's fee is returned if the task could not be executed. The stakes of honest worker nodes are also returned if the task was compromised by malicious actors. Each stake has an expiration time after which, if the holder has not been assigned to do work, the amount is fully returned. Offline nodes that are selected to perform work lose a part of their stake (e.g. 10\%).

\subsection{Tasks}
A client submits a task definition $T$ and a fee $F$ as a special transaction to the blockchain (\emph{PAY\_FOR\_TASK}), containing:
\begin{itemize}
	\item A description of the trained model: the model type (e.g.: multi-layer perceptron, convolutional neural network) and its architecture: the layers (e.g. dense, dropout, convolutional, batch normalisation, pooling) with the number of neurons, the activation functions (e.g. \emph{relu}, \emph{sigmoid}, \emph{tanh}), the loss function (e.g. softmax cross-entropy loss).
	\item The optimiser (e.g. SGD, Adam) and its  parameters (e.g. the initialiser, the learning rate).
	\item The stopping criterion (early-stopping, preset no. of epochs etc)
	\item The validation strategy (e.g. cross-fold validation, holdout) and the percentage of data from the training dataset set apart for validation. -- $D_{pct}^{(val)}$.
	\item Metrics of interest $K$ (e.g. accuracy, loss); the client will pay the best model based on these metrics on the test dataset.
	\item Dataset information: format (e.g. CSV, MNIST \cite{lecun-mnisthandwrittendigit-2010}), mini-batch size, training dataset percent out of the whole dataset ($D_{pct}^{(tr')}$), the hashes of at least 10 equally divided data batches from the original dataset $D$ and the size of the whole dataset $D$.
	\item Performance: expected training time ($t_{exp}$), hardware preferences (e.g. CPU vs GPU).
\end{itemize}

The fee $F$ is split between honest worker nodes and evaluators following a reward scheme $R$: a part is received by miners based on performance, a part by supervisors and another part by evaluators. All nodes are incentivised to participate in the PAI blockchain: clients receive a trained model, miners can receive block rewards and a fraction of the client's fee, supervisors and evaluators also receive a part of the client's fee. Miners compensate verifiers with a share of the block's reward.

\subsection{Protocol}
In our system, a client utilises the PAI blockchain to train a ML model and to pay for this service. After the client broadcasts the task to the PAI network, the miners and the supervisors are matched randomly by the network based on the worker nodes' preferences $P$ vs the task definition $T$.

The dataset $D$ is first split into:
\begin{itemize}
\item A training dataset, further revealed to the miners to perform ML work on it.
\item A validation dataset selected from the initial training dataset for ML validation.
\item A test dataset revealed to the evaluators when the final model should be tested.	
\end{itemize}

The training dataset is further split into equally-sized mini-batches. Typical sizes for a mini-batch range from 10 to 1000 records. A mini-batch is processed in each iteration. An epoch is a full training cycle, i.e. all mini-batches from the training dataset are processed.

Once training is started, miners iteratively improve their local models based on their work on mini-batches and based on messages they receive from fellow miners. Every miner shares the modifications to his/her model with other task participants. They also mine with several nonces obtained after each iteration.

The supervisors record all the messages during a task and look after malicious behaviours. Evaluators test the final models, select the best model for the client and distribute the client's fee.

Miners build blocks as in the Bitcoin protocol, but with the addition of useful work, i.e. honestly executing an iteration of the ML task. A miner scans the mempool, collects transactions, creates the block and adds additional information (extra fields) for the proof of useful work. In our PoUW, the nonce is a SHA-256 hash occupying 32 bytes.

To validate a block, verifiers will receive the input data, re-run the lucky iteration and check if the outputs are reproducible. A miner will send to verifiers all data and context necessary for the \emph{proof of useful work}.

\section{Workflow} \label{ML workflow}
 There are four ML task workflow stages: registration, initialisation, training and finalisation.
 
\subsection{Registration} \label{registration}
 In the registration phase, the client submits a PAY\_FOR\_TASK transaction. This transaction is included in a block called the \emph{task definition block}. Clients can revoke a task if it has not started yet by sending a REVOKE\_TASK transaction referencing the initial submission. The corresponding PAY\_FOR\_TASK transaction is thus annulled and the stake released. In case they want to adjust their bids, clients may also re-post their tasks before they start or expire. After updating, a task's maturity time is reinitialized.
 
 At least 2 miners and 3 supervisors are required to start a task. The number of supervisors can be in the interval $|S| \in [3, \max(3, sqrt |M|)]$, while the number of miners can be in the interval $[2, \omega_D * disk\_size(D) + \omega_t * t_{exp} + \omega_F * F]$, where $disk\_size$ is a function that returns the size of the dataset on the disk (in kB), $|S|$ is the number of the supervisors, $|M|$ is the number of miners, while $\omega_D, \omega_t$ and $\omega_F$ are network-wide coefficients. A task definition block not satisfying these requirements is deemed invalid.
 
 A worker node calculates if any of its live tickets is selected for participation in training. A ticket bought after task submission cannot participate in the training. For each ticket, the procedure is as follows:
 \begin{enumerate}
     \item calculate the cosine similarity between a ticket's preferences $P$ and the corresponding maximal subset of possible task preferences, $P_{max} \subseteq T$ ($P$ and $P_{max}$ are encoded as vectors with non-negative values):
     $s(P, P_{max})=\frac{\sum_{i=1}^n P_i {P_{max}}_i}{\sqrt{\sum_{i=1}^n P_i^2} \sqrt{\sum_{i=1}^n {P_{max}}_i^2}}$; $s(P, P_{max}) \in [0,1]$
     \item using a verifiable random function (VRF) \cite{MicaliRaVa99}, using the hash of the task definition block ($hash_{TDB}$) and the role on the ticket (supervisor, miner, etc) as inputs, produce a hash and a proof: $(hash, proof) = VRF_{sk}(hash_{TDB}||role)$ (see also \cite{Algorand}). Note: \textit{The hash looks random, but it is dependent on a secret key, $sk$, and the input string. Knowing the public key, $pk$, corresponding to $sk$ and the proof, one can verify if the hash was generated correctly.}
     \item if $s(P, P_{max}) \geq \omega_S \frac{hash}{2^{hashlen}-1}$, the ticket is selected to participate in task $T$. $\omega_S$ is a network-wide parameter, $hashlen$ is the bit-length of the hash.
 \end{enumerate}
 
 Selected worker nodes will send special transactions named \emph{applications} (\emph{JOIN\_TASK} transactions) that will be included in a subsequent block, called the \emph{participation block}. We seed the random number generator with the task definition block's hash to reproduce/verify the selection process.
 
\subsection{Initialisation}
\subsubsection{Key exchange and generation}
Worker nodes exchange their public keys to verify the received messages and the special transactions.

Supervisors use \emph{t-of-n threshold BLS signatures} to reach consensus, where $n=|S|$ and $t$ represents a 2/3 threshold ($t=\frac{2}{3}n$). Any subset of $t$ up to $n$ supervisors are able to sign a transaction and make it valid, but less than $t$ signers would render the transaction invalid. Supervisors run a modified version of the Joint-Feldman distributed key generation (DKG) protocol (\cite{Pedersen:1991:TCW:1754868.1754929}) to collectively generate their private BLS keys (see Supplementary material, \ref{dkg} and \ref{t_n}). Due to the BLS threshold signature properties, for any multi-party transaction $tx$, as soon as any $t$ signature shares are collected, a leader can reconstitute the global signature on the transaction and verify it as if the global private key (which is unknown) had been used for signing. At the end of the DKG protocol, all supervisors must post \textit{DKG\_SUCCESSFUL} transactions to the blockchain signed with their ECDSA secret keys containing their locally calculated t-of-n public key. If all supervisors post transactions with the same public key during a predefined time window, then the protocol is successful and the parties should proceed to the next phase. Otherwise, faulty supervisors are replaced, their stakes confiscated and the DKG protocol is restarted.

\subsubsection{Data preparation} \label{data prep}
Before submitting a task, the client privately splits the dataset $D$ into $p \geq 10$ consecutive fragments ($d_1, d_2 .. d_p$), such that $|d_1|=|d_2|=..=|d_{p-1}|;$ and $ |d_p|=|D|-(p-1)|d_1| < |d_1|$, then hashes the fragments, gets $H(d_1), H(d_2) .. H(d_p)$ and appends the hashes to the dataset section in the task definition. Hashes are public, but $D$ is known only to the client.

During the initialisation phase, the client and the worker nodes use the task definition block hash as a random seed to permute the hashes from $D$. The first $D_{pct}^{(tr')}$ \% of hashes will correspond to the initial training dataset and the client will reveal it to the worker nodes, while the rest will correspond to the test dataset, which is kept secret until the ML task is finished.

The \emph{validation dataset} is independently and deterministically derived by the miners from the initial training dataset based on the validation strategy. For example, in the case of \emph{holdout}, the validation dataset $D^{(val)}$ contains the last $D_{pct}^{(val)}$ \% mini-batches of the initial training dataset $D^{(tr')}$. The final training dataset is obtained by subtracting the validation dataset from the initial training dataset: $D^{(tr)}= D^{(tr')} \setminus D^{(val)}$.

The final training dataset is further split into $m$ mini-batches $b_1...b_m; |b_1|=|b_2|=..=|b_{m-1}|$ and $|b_m|=|D^{(tr)}|-(m-1)|b_1| \leq |b_1|$ (size specified by the client in the task preferences). We use one of the following two methods to assign batches to miners:
\begin{description}\label{mini-batch assignment}
    \item[Consistent hashing with bounded loads], a method inspired by \cite{Karger:1997:CHR:258533.258660} and \cite{Mirrokni2018}, in which one computes the hashes of mini-batches concatenated with the current epoch number $\xi$: $H(b_1||\xi)..H(b_n||\xi)$. Then, maps every hash to a \emph{consistent hash ring} of size $2^{\kappa}$ using the modulo function: $H \bmod 2^\kappa$, where $\kappa$ is the exponent of the mapping space (chosen network-wide, e.g. $\kappa=10$). Each miner should process the mini-batches with the hashes between his/her hash and his/her successor's hash on the hash ring. To prevent discrepancies between the amount of assigned work, we never let a miner process more than $c m /|M|$ mini-batches, where $1 \leq c \leq 2$ is a public constant. We do so by assigning the current mini-batch to the next miner until the above condition is met.
    \item[Interleaved parts], a procedure in which miners and mini-batches are sorted lexicographically by their IDs ($M=\{M_1..M_n\}$) and their hashes, respectively. The allocation procedure specifies that the first miner gets the first mini-batch, the second one gets the second and so on, until all $|M|$ miners are assigned to the first $|M|$ mini-batches. Then, the first miner receives the $|M|+1$-th mini-batch and so on, until all $m$ mini-batches are assigned. For each epoch $\xi$, a miner $M_i$ is first assigned to mini-batch $b_{i+\xi-1}$, then the assignment continues with $b_{i+\xi-1+|M|}$, $b_{i+\xi-1+2|M|}...$ in a rotating manner to also include the first initially uncovered positions $[1..i+\xi]$ (the mini-batches are seen as a circular data structure).
\end{description}

If a miner quits or a new one joins a task, the above assignment is re-evaluated. Using a random number generator seed based on the hash of the task definition block, we ensure that the miners know in advance which mini-batches should be processed and the order. The steps can also be reproduced during the verification.

\subsubsection{Data storage}
Supervisors store the data during the ML training. They can use one of the following redundancy schemes:
\begin{description}
    \item[Full replicas] Every supervisor keeps a copy of $D^{(tr')}$. This has a high redundancy factor (\cite[p.~36]{Duminuco2009}) equal to the number of supervisors $\beta = \frac{|data_{red}|}{|data|}=|S|$, and a miner can selectively download mini-batches.
    \item[Reed-Solomon]The supervisors store $D^{(tr')}$ using a \emph{$(k+h)$ Reed-Solomon storage scheme} (\cite{ReedSolomon}). They split the training dataset into $k+h$ equal fragments, where the total number of the supervisors is roughly $|S| \approx k+h$ and $k \approx [1/3..2/3]*|S|$. To get the full dataset, a miner should download data from $k$ supervisors and re-construct it. The redundancy is low $\beta = \frac{k+h}{k}=[1.5..3]$, but miners perform extra computation to re-create the dataset.
\end{description}

\subsection{Training}
Miners start working on a ML task by initialising the weights and biases ($\boldsymbol{\theta}$) of their local models. At each iteration they fetch a mini-batch, perform \emph{stochastic gradient descent} (SGD) on it (\cite[p. 147]{Goodfellow2016}), then communicate what they changed in $\boldsymbol{\theta}$ (weight updates) to the other worker nodes. Cost functions in most ML algorithms are sums of loss functions $L$ over training records and their global gradient is an expectation. To minimise the global loss, instead of using all records, one could sample a mini-batch $b_i$ and calculate a partial gradient $\mathbf{g^{(i)}}$ that approximates the global gradient: $\mathbf{g^{(i)}}=\frac{1}{|b_i|}\nabla \sum_{j=1}^{|b_i|}L(x^{(j)}, y^{(j)}, \theta)$, where $x$ and $y$ represent the features and targets per record. Using this gradient, the classical SGD algorithm updates the model as follows: $\theta \leftarrow \theta - \epsilon \mathbf{g^{(i)}}$, where $\epsilon$ is the learning rate.

The size of a gradient is equal to the size of the local model, so it is impractical for a miner to send all the gradient updates to the network. To overcome this issue, we use the "dead-reckoning" scheme from \cite{Strom2015}, in which we update only weights that under/over-flow a certain $\tau$ value and communicate the coordinates of these updates in the local model. Gradients under/over $\tau$ accumulate and are applied later. Learning is not negatively impacted by this delay.

\begin{figure}[!t]
    \centering
    \includegraphics[width=0.48\textwidth]{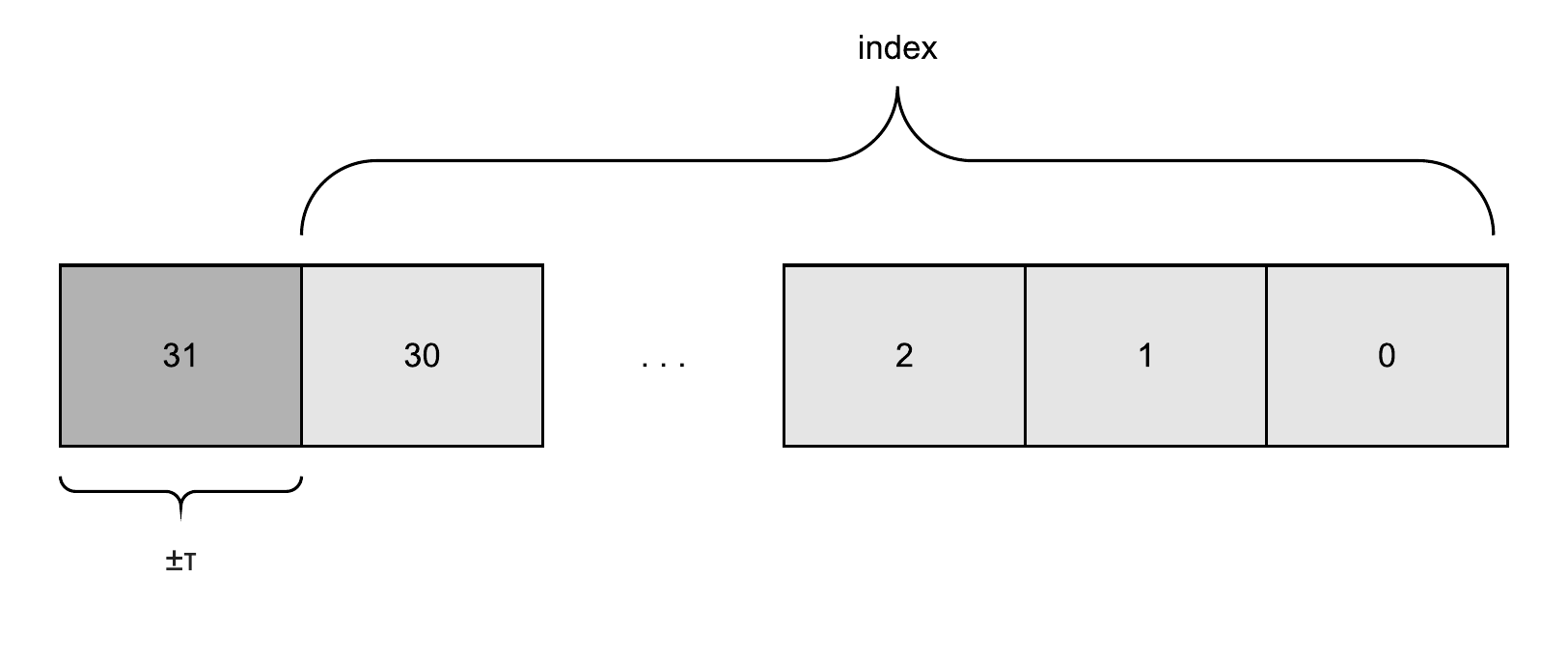}
    \caption{\textbf{Message map.} Leftmost bit is for $\tau$, the rest is for the index.}
    \label{fig:mmap}
\end{figure}

An iteration consists of the steps described in \textbf{Algorithm \ref{alg:nodeop}}, adapted from \cite{Strom2015}. See Table \ref{tab:notation} for notation.
	\begin{itemize}[leftmargin=*]
	 \item The miner fetches a mini-batch ($b_i$) according to a method from subsection \ref{mini-batch assignment}.
	 \item Updates the local model with weight updates received from peers.
     \item Derives local gradients using backpropagation and adds them to a residual gradient.
     \item Prepares a message map, a list $\boldsymbol{\delta^{(\ell)}}$ containing the addresses (indices) of the gradients whose values exceed a threshold value $\pm \tau$ in the residual gradient, where $\tau$ is given by the client. An element $e({0|1}, i)$ of the list has 4 bytes. The leftmost bit is 0 when the gradient residual at index $i$ is less than $-\tau$ and 1 when it is greater than $+\tau$). The other 31 bits are for the index $i$ (unsigned integer from 0 to $2^{31}-1$) (see Fig. \ref{fig:mmap}).
     \item Applies the weight updates from the message map to the local replica of the DNN.
     \item Receives and uncompresses peers' messages.
     \item Communicates to the network a message containing: version of current implementation, ML task ID, message type (\textit{IT\_RES}), epoch -- $\xi$, number of peer updates, message map of weight updates -- $\delta^{(\ell)}$, values of metrics -- $K^{(tr)}$, start and finish times -- $t^{(s)}$, $t^{(f)}$, hash of mini-batch, hash of the list with peer messages applied at current iteration, hash of initial model state $\theta$, hash of initial gradient residual, hash of zero-nonce block intended to be mined $k$ iterations in the future -- $hash(ZNB)$, hash of list with uncompressed peer messages to be used in the next iteration (received in step 20) and the DER-encoded ECDSA signature of the miner (see Table \ref{tab:serialisation}).
     \item Mines with several nonces.
   \end{itemize}

\begin{algorithm}
\caption{Work done during an iteration}\label{alg:nodeop}
\begin{algorithmic}[1]
\Procedure{Iteration()}{}
\State $\mathbf{X^{(tr)},y^{(tr)}} \leftarrow Load(b_i)$ \Comment{Load a mini-batch}
\ForEach {$\boldsymbol{\Delta^{(p)}_i} \in \boldsymbol{\Delta^{(p)}}$}                   
	\State $\boldsymbol{\theta'} \leftarrow \boldsymbol{\theta} - \epsilon \boldsymbol{\Delta^{(p)}_i}$ \Comment{Update local replica with peer values}
\EndFor
\State $\boldsymbol{g^{(\ell)}} \leftarrow Backprop()$\Comment{Compute local gradients using backpropagation}
\State $\mathbf{g^{(r)'}} \leftarrow \mathbf{g^{(r)}} + \boldsymbol{g^{(\ell)}}$ \Comment{Update gradient residual}
\State $\mathbf{g^{(r)''}} \leftarrow \mathbf{g^{(r)'}};\boldsymbol{\Delta^{(\ell)}} \leftarrow \boldsymbol{0_{n}}; \boldsymbol{\delta^{(\ell)}} \leftarrow {} $ \Comment{Initialise final residual gradient, local weights list and message map ($n$ -- length of the gradient)}
\ForEach {$g^{(r)'}_i \in \mathbf{g^{(r)'}}$}
\If{$g^{(r)'}_i > +\tau$}
\State $\boldsymbol{\delta^{(\ell)}} \leftarrow \boldsymbol{\delta^{(\ell)}} \cup e(\mathbf{1}, i)$
\State $\Delta^{(\ell)}_i \leftarrow +\tau$
\State $g^{(r)''}_i \leftarrow g^{(r)'}_i - \tau$
\ElsIf{$g^{(r)'}_i < -\tau$}
\State $\boldsymbol{\delta^{(\ell)}} \leftarrow \boldsymbol{\delta^{(\ell)}} \cup e(\mathbf{0}, i)$
\State $\Delta^{(\ell)}_i \leftarrow -\tau$
\State $g^{(r)''}_i \leftarrow g^{(r)'}_i + \tau$
\EndIf
\EndFor
\State $\boldsymbol{\theta''} \leftarrow \boldsymbol{\theta'} - \epsilon \boldsymbol{\Delta^{(\ell)}}$ \Comment{Update local replica with local values}
\State $\boldsymbol{K^{(tr)}} \leftarrow FwdPass(\boldsymbol\theta'', \mathbf{X^{(tr)}, y^{(tr)}})$ \Comment{Evaluate metrics on the training mini-batch}
\State $\boldsymbol{\delta^{(p)}} \leftarrow $ \emph{Get peers message maps()} \Comment{Receive, uncompress peer messages and extract the message maps.}
\State $ \boldsymbol{\Delta^{(p)}} \leftarrow$ \emph{Rebuild gradients}($\boldsymbol{\delta^{(p)}}$)
\State $msg \leftarrow$ \emph{Build message} (serialise the variables in Table \ref{tab:serialisation} in an IT\_RES message, sign and compress)
\State $Send(msg)$ \Comment{Broadcast message to network} 
\State $Mine(...)$ \Comment{Mine}
\EndProcedure
\end{algorithmic}
\end{algorithm}

\begin{table}
\renewcommand{\arraystretch}{1.3}
\caption{Payload of an IT\_RES message.}
\label{tab:serialisation}
\centering
\begin{tabular}{@{}|l|l|l|l|@{}}
\toprule
\rowcolor[HTML]{C0C0C0} 
	field & bytes & type & notes \\ \midrule
	$version$ &  2   &   short & message version   \\ \midrule
	$task\_id$ &  16   &   uuid & task id   \\ \midrule
	$msg\_type$ &  1  &   char & IT\_RES   \\ \midrule
	$\xi$ &   2   &   ushort &  epoch \\ \midrule
	$|\delta^{(\ell)}|$ &   2   &   ushort &  no. of peer updates \\ \midrule
  	$\delta^{(\ell)}$   &  $4*|\delta^{(\ell)}|$  & list   &  see Fig. \ref{fig:mmap}  \\ \midrule
   $|K^{(tr)}|$ &   2   &   ushort & no. of metrics   \\ \midrule
   $K^{(tr)}$ &   4 * $|K^{(tr)}|$   &   list & metrics   \\ \midrule
   $t^{(s)}$ &   4   & uint &  UNIX timestamp   \\ \midrule
   $t^{(f)}$ &   4   & uint &  UNIX timestamp   \\ \midrule
   $hash(b_i)$  &   32   & uint256 &   SHA256    \\ \midrule
   $hash(\Delta^{(p)})$     &   32   & uint256 &   SHA256    \\ \midrule
   $hash(\theta)$  &   32   & uint256 &   SHA256    \\ \midrule
   $hash(g^{(r)})$  &   32   & uint256 &   SHA256    \\ \midrule
   $hash(ZNB)$  &   32   & uint256 &   SHA256    \\ \midrule
   $hash(\delta^{(p)})$     &   32   & uint256 &   SHA256    \\ \midrule
   $sgn$ &  65   &  string & DER-encoded ECDSA     \\ \bottomrule
\end{tabular}
\end{table}

\begin{table}
\renewcommand{\arraystretch}{1.3}
\caption{Notation used in Algorithm \ref{alg:nodeop}.}
\label{tab:notation}
\centering
\begin{tabular}{@{}|l|l|@{}}
\toprule
\rowcolor[HTML]{C0C0C0} 
	symbol & meaning \\ \midrule
	$\mathbf{X^{(tr)}}$ &  features in the mini-batch   \\ \midrule
	$y^{(tr)}$ &  targets in the mini-batch   \\ \midrule
	$\theta$ &  model's state (weights and biases)\\ & at the beginning of iteration   \\ \midrule
	$\theta'$ &  model's state (weights and biases)\\ & after applying peer updates   \\ \midrule
	$\theta''$ &  model's state (weights and biases)\\ & after applying local updates   \\ \midrule
	$\epsilon$ &  learning rate \\ \midrule
	$\boldsymbol{\Delta^{(p)}}$ &  weight updates from peers   \\ \midrule
	$\boldsymbol{\Delta^{(\ell)}}$ &  local weight updates   \\ \midrule
	$\boldsymbol{g^{(\ell)}}$ &  local gradient   \\ \midrule
	$\mathbf{g^{(r)}}$, $\mathbf{g^{(r)'}}$, $\mathbf{g^{(r)''}}$ &  residual gradient at different steps  \\
	 \midrule
	$\boldsymbol{\delta^{(\ell)}}$ &  message map  \\ \midrule
	$\boldsymbol{\delta^{(p)}}$ &  message maps from peers   \\ \midrule
	$K^{(tr)}$  & metrics on mini-batch   \\ \midrule
	$t^{(s)}$  &  iteration start time   \\ \midrule
	$t^{(f)}$  &  iteration end time \\ \bottomrule
\end{tabular}
\end{table}

If a miner finishes earlier the work for an epoch, then he/she will wait for the majority of his/her peers to also complete and will apply their peer updates \emph{ad interim}.

For each epoch $\xi$, a leader supervisor is chosen in a round-robin fashion using the formula: $(i-1) \mod \xi + 1$, where $i$ is the rank of the supervisor's ID after sorting all supervisors in lexicographical order. If the next potential leader (index $(i-1) \mod \xi + 2$) submits a valid multi-party \emph{REPLACE\_LEADER} special transaction backed by at least $2/3$ of supervisors then the current leader is replaced by the next one. The transaction contains the reason why the leader was replaced (offline, too slow, malicious etc). This procedure can continue until a good leader is found.

Supervisors help miners to prove that they performed honest ML work by recording the messages sent during the training, the \emph{message history}. Verifiers are only interested in the segments of the message history that can demonstrate useful work. Therefore, during an epoch the leader will pack sets of consecutive messages into \emph{slots}. Supervisors agree on the exact order of the messages in a slot using preferential voting (e.g. Schulze method \cite{Schulze2011}). After the votes are cast, the leader computes the result per slot and publishes the hash of it as a \emph{MESSAGE\_HISTORY} transaction. The slot raw data is kept by all supervisors and can be downloaded by miners or verifiers.

Worker nodes lose their stake if they skip assigned iterations (5\% for miners and 10\% for supervisors). A worker node rejoining a task has to catch up with the latest weight updates. Supervisors can add another miner to the group if the number of miners drops below 80\% of the initial size. A supervisor is replaced when he/she fails to record more than 10\% of the iterations. The leader initiates the replacement procedure by publishing a special transaction called \emph{RECRUIT\_WORKER\_NODE}. We provide all the details about the detection of offline nodes and the replacement procedure in the Supplementary material (\ref{Crash}). 

\subsection{Finalisation}
The client provides the test dataset to the evaluators and they will verify if the client did not cheat (i.e. the hashes match the withheld test mini-batches). They check the model from each miner on the test dataset and they send the model with the best performance to the client. Evaluators are selected using a matching algorithm similar to the one used for worker nodes.

Evaluators must produce identical conclusions containing ML metrics. A conclusion is published using the special \emph{CHARGE\_FOR\_TASK} transaction. If less than 2/3 of the evaluators produce identical reports, subsequent rounds of evaluators are drawn until a 2/3 consensus is reached. An evaluator can cheat by waiting to see the conclusions of others appearing in the mempool, and then publish an identical one. To prevent this, we break the procedure in two phases: 1) all evaluators post conclusions to the mempool in an encrypted form; 2) once they see that all conclusions have been posted, each of them then posts the decryption key. We call this method \emph{commit-then-reveal}.

\begin{figure}[!t]
    \centering
    \includegraphics[width=0.48\textwidth]{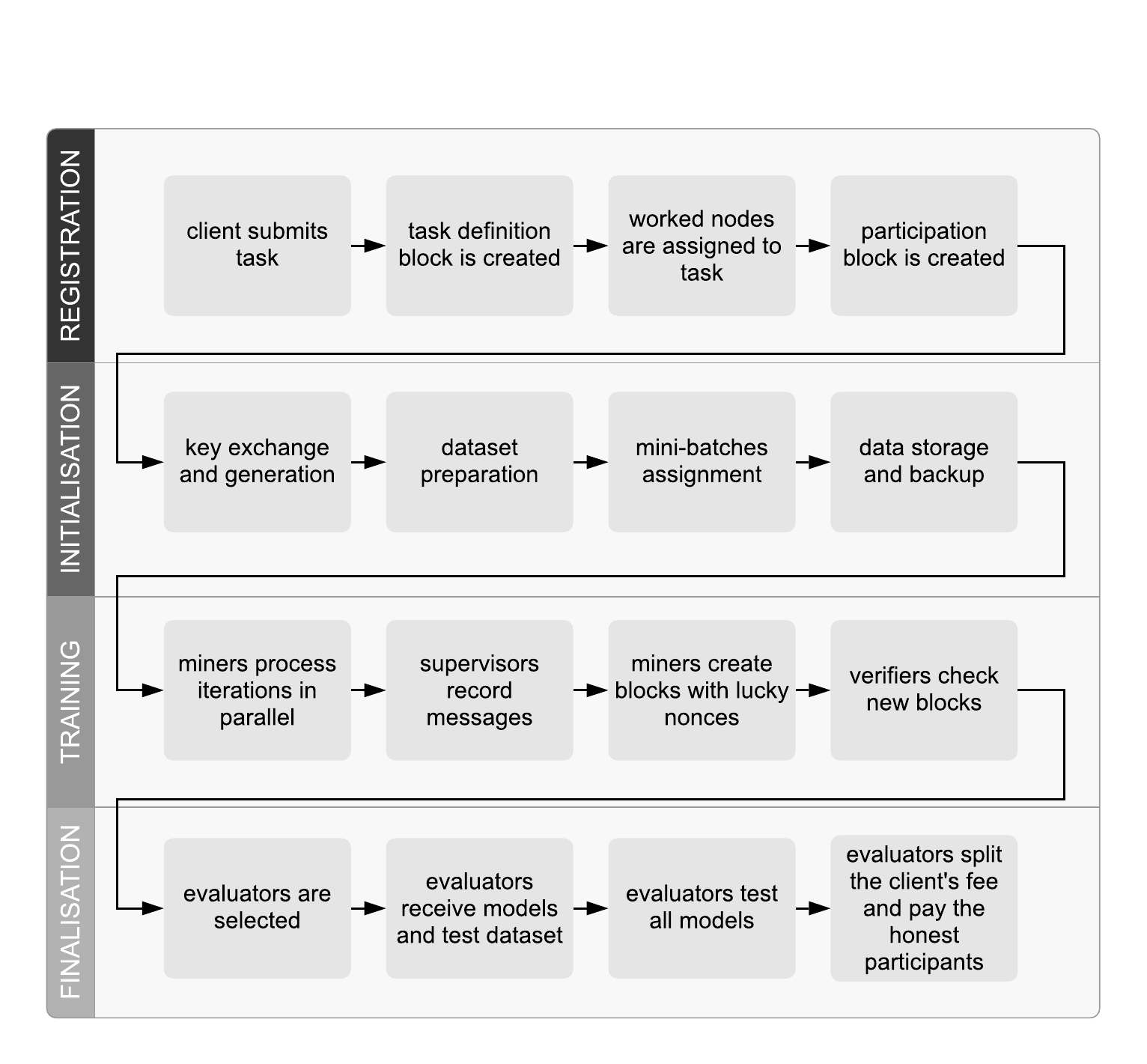}
    \caption{\textbf{ML task steps.} Summary of main events during a typical ML training procedure in the PAI blockchain.}
    \label{fig:ml_steps}
\end{figure}

We summarised all the above steps in Fig. \ref{fig:ml_steps}.

\section{Proof of Useful Work} \label{PoUW}
\subsection{Mining} \label{Mining}
After each iteration, a miner has the right to mine a block. In the classical Bitcoin, a miner can obtain different hashes of the block header by varying the nonce. We limit the number of nonces to $a= \omega_B * disk\_size(b_i) + \omega_M * model\_size(\boldsymbol{\theta''})$, where $model\_size$ is a function returning the number of weighs and biases in the model, while $\omega_B$ and $\omega_M$ are network-wide coefficients. We want to ensure that hashing is insignificant and that most computing power is spent for ML training.

\begin{figure}[!t]
    \centering
    \includegraphics[width=0.48\textwidth]{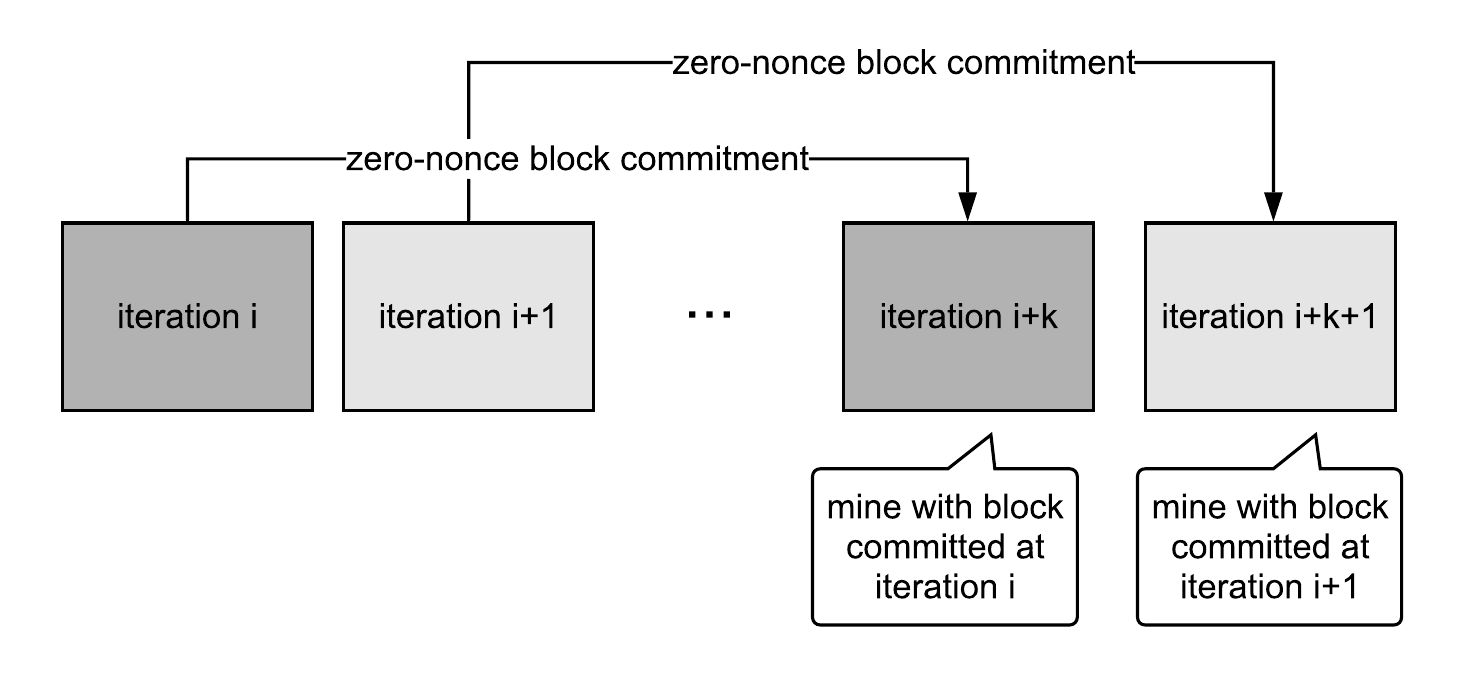}
    \caption{\textbf{Zero-nonce block commitment.} A miner announces $k$ iterations in advance the zero-nonce block to be mined.}
    \label{fig:zbc}
\end{figure}

A miner might also generate billions of hashes by manipulating the transactions included in the block (e.g. changing the timestamp). Therefore, we require that $k$ iterations before, the miner should "commit" to a \emph{zero-nonce block} (ZNB), as illustrated in Fig. \ref{fig:zbc}. A ZNB is built using a fixed header and a fixed set of transactions, with the nonce and other auxiliary fields necessary for PoUW set to zero. A miner should include the ZNB hash in the IT\_RES message at iteration $i$. At iteration $i+k$, the miner replaces the nonce from the ZNB with the nonces obtained with by-products of ML training, as described in Algorithm \ref{alg:mining}. The nonce is a double hash of the concatenation of the local model at the end of the iteration ($\theta''$) and the local gradients. We call the first hash obtained with the by-products of ML work the \emph{nonce precursor}.

\begin{algorithm}
\caption{Mining operation}\label{alg:mining}
\begin{algorithmic}[1]
\Procedure{Mine(...)}{}

\State $nonce_{precursor} \leftarrow hash(\boldsymbol{\theta''}|\Delta^{(\ell)})$ \Comment{Build nonce precursor}
\State $nonce \leftarrow hash(nonce_{precursor})$ \Comment{Build nonce}
\State $a= \omega_B * disk\_size(b_i) + \omega_M * model\_size(\boldsymbol{\theta''})$ \Comment{No. of nonces allowed}
\For{$j \gets 0$ to $a-1$}                    
	\State $success = MineWithNonce(nonce + j)$
	\If{$success$}
\State $Store(\boldsymbol{\theta}, b_i, g^{(r)}, \Delta^{(p)}, h_i)$ \Comment Miner stores: model from beginning of iteration, mini-batch, gradient residual, peer updates and the relevant message history IDs
\State $block \leftarrow MakeBlock(...nonce, hash(h_i),$ $ hash(msg)...)$ \Comment{New block with PoUW fields.}
\State $AddToPAIBlockChain(block)$ \Comment{Adds the block to the PAI blockchain.}
\EndIf
 \EndFor
\EndProcedure
\end{algorithmic}
\end{algorithm}

\begin{figure}[!t]
    \centering
    \includegraphics[width=0.4\textwidth]{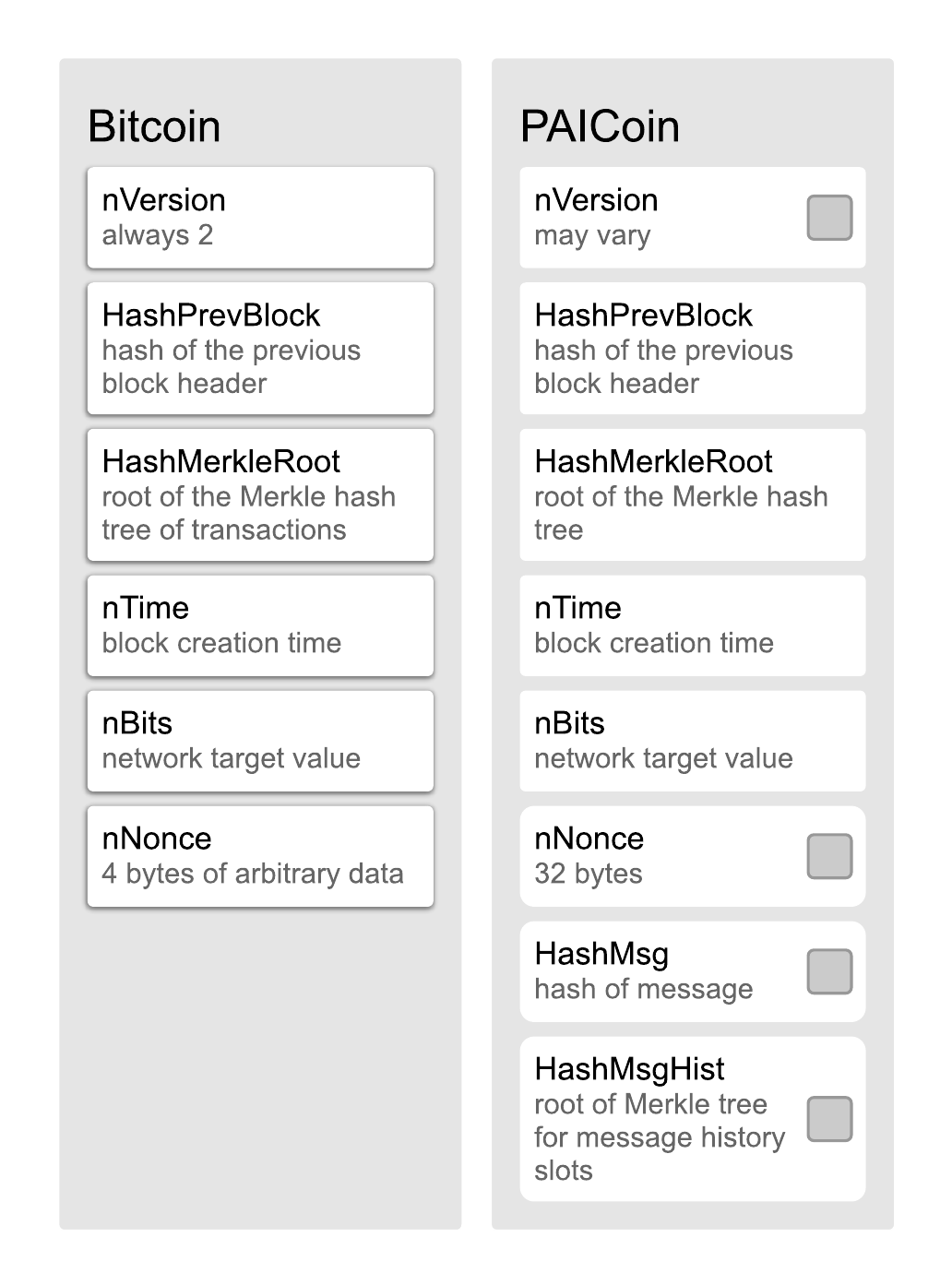}
    \caption{\textbf{Block header differences between Bitcoin and PAICoin.} Differing fields are marked by a square sign. We used the same notation and block structure as in \cite{bitcoindeveloper}.}
    \label{fig:block-header}
\end{figure}

Upon successful mining, the miner stores the initial model state (weights and biases), the mini-batch, the initial gradient residual and the peer updates, but also downloads and stores the relevant message history slots from the supervisors. Furthermore, creates a new block with $hash(h_i)$ and $hash(msg)$ as extra fields for lookup, where $hash(h_i)$ refers to the Merkle tree of the message history slots containing the IT\_RES messages from iteration $i$ up to iteration $i+k$, while $hash(msg)$ is the hash of the IT\_RES message corresponding to the lucky iteration. After successful verification, the block is added to the PAI blockchain. Fig. \ref{fig:block-header} shows the block header particularities of PoUW vs. the Bitcoin.

For settlement, transactions are added from the mempool to the blockchain through mining. When there are periods without client tasks, to ensure continuous mining, miners will work on several network-wide predefined tasks, such as protein structure prediction using DNN. Miners would not get a client's fee, only the block's reward and the transactions' fees. Exceptionally, supervisors, verifiers and evaluators would get paid from the block's reward. After all blocks with non-zero subsidy have been mined, they will be paid from the transactions' fees. To prevent a shortage of miners, the network should have a limited number of dedicated mining agents belonging to Project PAI.

\subsection{Verification}\label{Verification}
Verification of a mined block is delegated to the \emph{verifier} because re-running an iteration is computationally expensive. Verification has the following steps:
\begin{itemize}[leftmargin=*]
\item Ten \emph{verifiers} are automatically selected based on the mined block hash (as a random seed).
\item Verifiers first check if the hash of the block is under the network target T (as in Bitcoin).
\item They extract the fields from the block: $nonce, hash(h_i)$ and $hash(msg)$ and receive the mini-batch, the local model from the start of the iteration, the gradient residual, the peers' weight updates and the relevant message history part.
\item After extracting information from the message history, they can check if the miner is legit and if the ticket and the ML task are valid.
\item A verifier compares the hashes and decides if the message history slot was registered as a special transaction in a previous block and if the hash of the starting model, the gradient residual and the peers' updates were announced in the previous iteration.
\item They verify if batch $b_i$ exists and if it should have been processed at that particular time in that order, as given by the preset order of messages from the initial allocation.
\item From message history, verifiers can check if the $i+k$ commitment is valid, by comparing it with the zero-nonce block version of the currently mined block.
\item Given the mini-batch, the starting local model, the peers' updates and the gradient residual, verifiers re-run steps 2-18 from the Algorithm \ref{alg:nodeop} and check if the obtained metrics coincide with the ones sent to the network.
\item Verifiers compress, hash the peer messages and check if the results match the hash reported in the previous message.
\item With the end iteration model state and the local weights map calculated, a verifier will reconstruct the nonce precursor and the nonce and compare it with the one provided in the mined block.
\item Verifiers will also check if the metrics improved over a time window $\Delta t$.
\item Each verifier will post an encrypted digest (as a transaction) containing the nonce precursor to the mempool.
\item The miner watches for digests to appear and decides at some point to post a \emph{COLLECT\_VERIFICATIONS} special transaction. The miner is interested to get as many confirmations as possible because the next block subsidy is proportional to them (in increments of 10\%, up to 100\% of the usual Bitcoin subsidy per block when 10 approvals are collected) and he/she wants other miners to build on top of his block in the future.
\item Verifiers will reveal their digests by providing the encryption key as another transaction. The digests also contain the metrics reported on the task.
\end{itemize}

Regular peer nodes check if the block hash is under the network target as in the Bitcoin protocol, but they also watch for verifiers' digests to verify if the nonce was generated from the nonce precursor. The transactions containing digests for the previous block are included in the current block to prove the subsidy amount in the \emph{coinbase transaction} (the transaction that pays the miner for finding the block). 

\section{Implementation} \label{Implementation}
\begin{figure}[!t]
\centering
\includegraphics[width=0.48\textwidth]{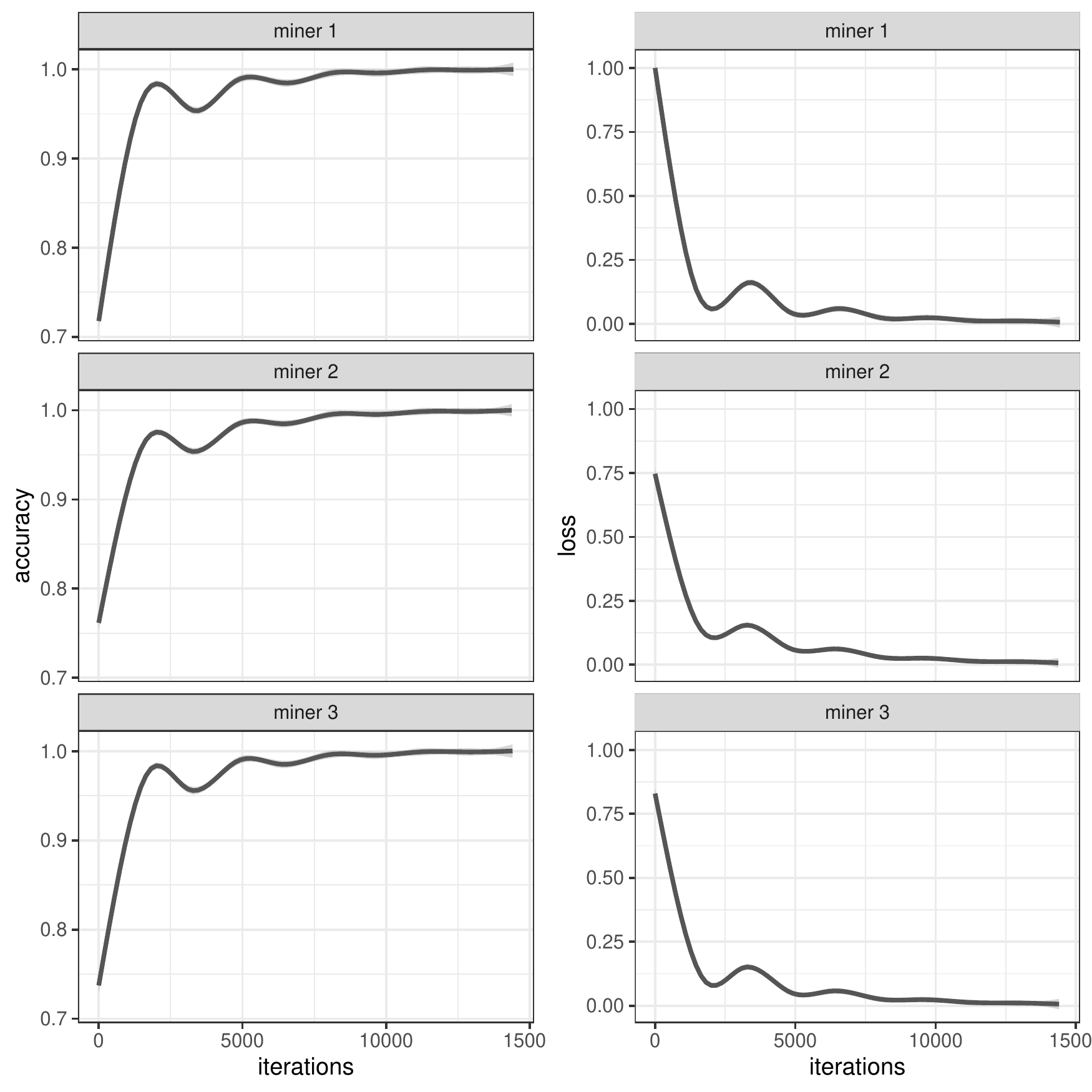}
\caption{\textbf{Convergence.} In this example, a distributed ML system with 3 miners improves the accuracy and minimises the loss as the training is progressing.}
\label{fig:convergence}
\end{figure}

We implemented an early stage proof-of-concept (PoC) of PoUW. Our work can be found at \def\UrlFont{\bfseries}\url{pouw.projectpai.com }. There are three software repositories and all projects have detailed step-by-step setup and usage instructions.:
\begin{description}
\item[PoUW Core] contains the code related to the distributed ML training (Algorithm \ref{alg:nodeop}), verification (Subsection \ref{Verification}) and mining (see Algorithm \ref{alg:mining}). For the ML training, we used MXNet \cite{Chen2015}, which is suited for high performance machine learning and has support for dynamic computational graphs. According to our preliminary tests, MXNet is generally faster than several other similar ML frameworks. MXNet is also used by Amazon in their ML cloud offerings and it is an open-source project, part of the Apache Incubator.
\item[PoUW Blockchain] is the underlying blockchain used for our PoC and it is a modified version of Project PAI's blockchain. It contains the block header modifications, special transactions and other blockchain logic. PoUW Core cannot function without PoUW Blockchain and viceversa.
\item[PoUW Simulation] is a project based on Kubernetes \cite{Kubernetes} to simulate a PoUW environment with different actors for how to setup a simulation). It can be used to study a PAI network on a local or cloud machine. Fig. \ref{fig:convergence} shows an example of how our ML system is improving the metrics during the training. To collect these metrics, 3 miners trained a fully-connected DNN using the standard MNIST dataset from \cite{lecun-mnisthandwrittendigit-2010} on a NC6 Microsoft Azure machine (Intel Xeon E5-2690v3 CPU, 56 GB RAM and 1 x K80 GPU). The convergence topic and how the distributed system behaves with different numbers of training nodes on various data sets are outside the scope of this paper (see \cite{Strom2015} for more information on these matters).
\end{description}
PoUW is now part of the Project PAI initiative. More details about the project are available on the parent page at \url{projectpai.com}.

\section{Discussion} \label{Discussion}
Our protocol brings several advantages over the classical Bitcoin. We use a modified hybrid PoW/PoS consensus with better security and we provide more rewards to network participants. However, due to its inherent complexity, our solution takes into account and mitigates potential performance and security risks. In the following subsections, we provide an economical analysis and discuss about several performance optimisations and adversarial scenarios important for the security of the system.

\subsection{Economical analysis}

Our PoUW solution is more profitable than Bitcoin mining and it is cheaper than ML cloud solutions. To prove it, we compare the costs of a client using a cloud ML solution vs. ours, and the return on investment (ROI) for miners in two scenarios: Bitcoin and PoUW mining. We assume that a client will not pay an hourly fee higher than what cloud providers are charging for a ML-capable virtual machine. Miners would not participate if their investment and operating costs exceed their profits. Miners would switch to Bitcoin mining if that would be more profitable.

To study the ROI and the profitability, we published an online PoUW cost calculator available online on the project page under the 'ROI Calculator' link. The variables' names and formulas are explained in the online document. We briefly describe the calculator:
\begin{itemize}[leftmargin=*]
    \item The first sheet contains global parameters, such as the client's hourly fee (in USD), the average number of paid participants per task, the ML distributed system efficiency, the price of the PAICoin and the Bitcoin, the electricity price, no. of active miners in the PAI network and the PAI block revenue.
    \item The second sheet is a price study containing profits for the most popular Bitcoin mining rigs.
    \item The third sheet contains a comparison between using cloud ML training solutions from Amazon, Microsoft and Google vs. four local deep learning workstation configurations (two hardware configurations from \cite{deep_learning_workstations} and two from Exxact \cite{exxact}).
    \item The fourth sheet shows the client's profit if he/she uses our system.
\end{itemize}

\begin{figure}[!t]
\centering
\includegraphics[width=0.48\textwidth]{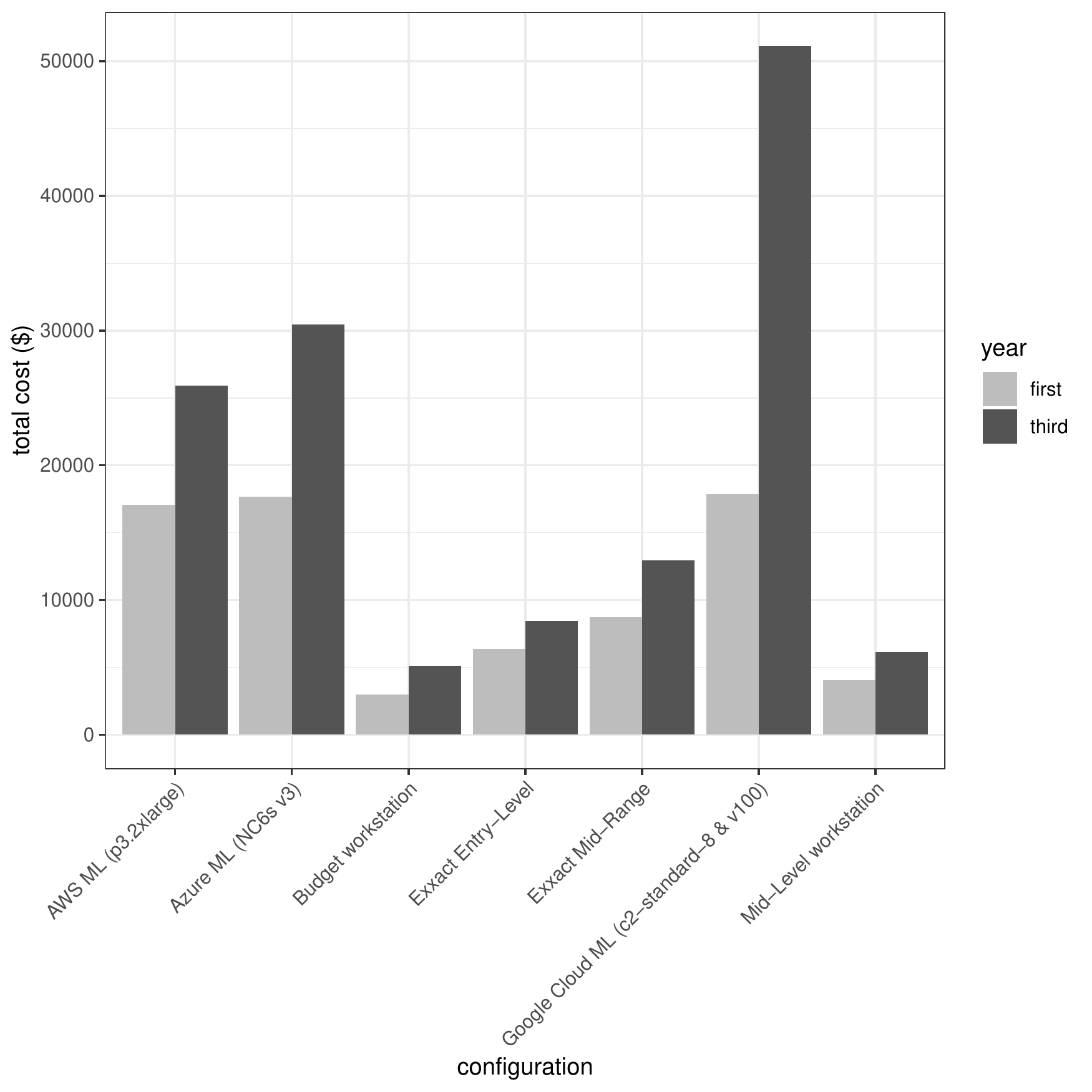}
\caption{\textbf{Total costs at 1 and 3 years.} Budget, Mid-Level and Exxact configurations are local solutions. AWS, Azure and Google are cloud solutions.}
\label{fig:total_costs}
\end{figure}

We assume a PoUW miner invests initial capital to buy ML hardware that is amortised in 3 years. It is cheaper for a miner to buy and use a local workstation than use similar cloud machine configurations. The total costs after 1 and 3 years for local vs. cloud scenarios are shown in Fig. \ref{fig:total_costs}. The formula for the \emph{return on investment} (ROI) as a percentage is:
\begin{eqnarray*}
R = 100 (\frac{g\left( \frac{F_h}{Q} + 6 \frac{W}{U}\right)}{\frac{C_{H}}{26280} + E} - 1),
\end{eqnarray*} where $g$ is the number of GPUs in the configuration, $F_h$ is the client hourly fee (\emph{fee\_usd\_client\_1h} from the cost calculator), $Q$ is the average number of paid participants per task (\emph{paid\_participants\_per\_task}), $W$ -- miner's block revenue (\emph{revenue\_pai\_block * price\_usd\_paicoin}), $U$ -- average number of miners in the network at any moment (\emph{miners\_active}), $C_H$ -- initial cost of hardware (\emph{cost\_usd\_hwd}), $E$ -- electricity price per hour (\emph{power\_kwh\_1h * cost\_usd\_1h}). To assess how ROI changes in respect to the input variables, we provide its partial derivatives:
\begin{align*}
& \frac{dR}{dF_h}=\frac{2628000 g}{Q \left( C_H + 26280 E \right)} \\
& \frac{dR}{dQ}=-\frac{2628000 g \cdot F_h}{Q^2 \left( C_H + 26280 E \right)} \\
& \frac{dR}{dW}=\frac{15768000 g}{U\left(C_H + 26280 E \right)} \\
& \frac{dR}{dU}= - \frac{15768000 g \cdot W}{U^2\left(C_H + 26280 E \right)} \\
& \frac{dR}{dC_H}= - \frac{2628000 g \cdot \left(F_h \cdot U + 6 \cdot Q \cdot W \right)}{Q U\left(C_H + 26280 E \right)^2} \\
& \frac{dR}{dE}= -\frac{69063840000 g \left(F_h \cdot U + 6 \cdot Q \cdot W \right)}{QU\left(C_H + 26280 \cdot E \right)^2}
\end{align*} 
        
\begin{figure}[!t]
\centering
\includegraphics[width=0.48\textwidth]{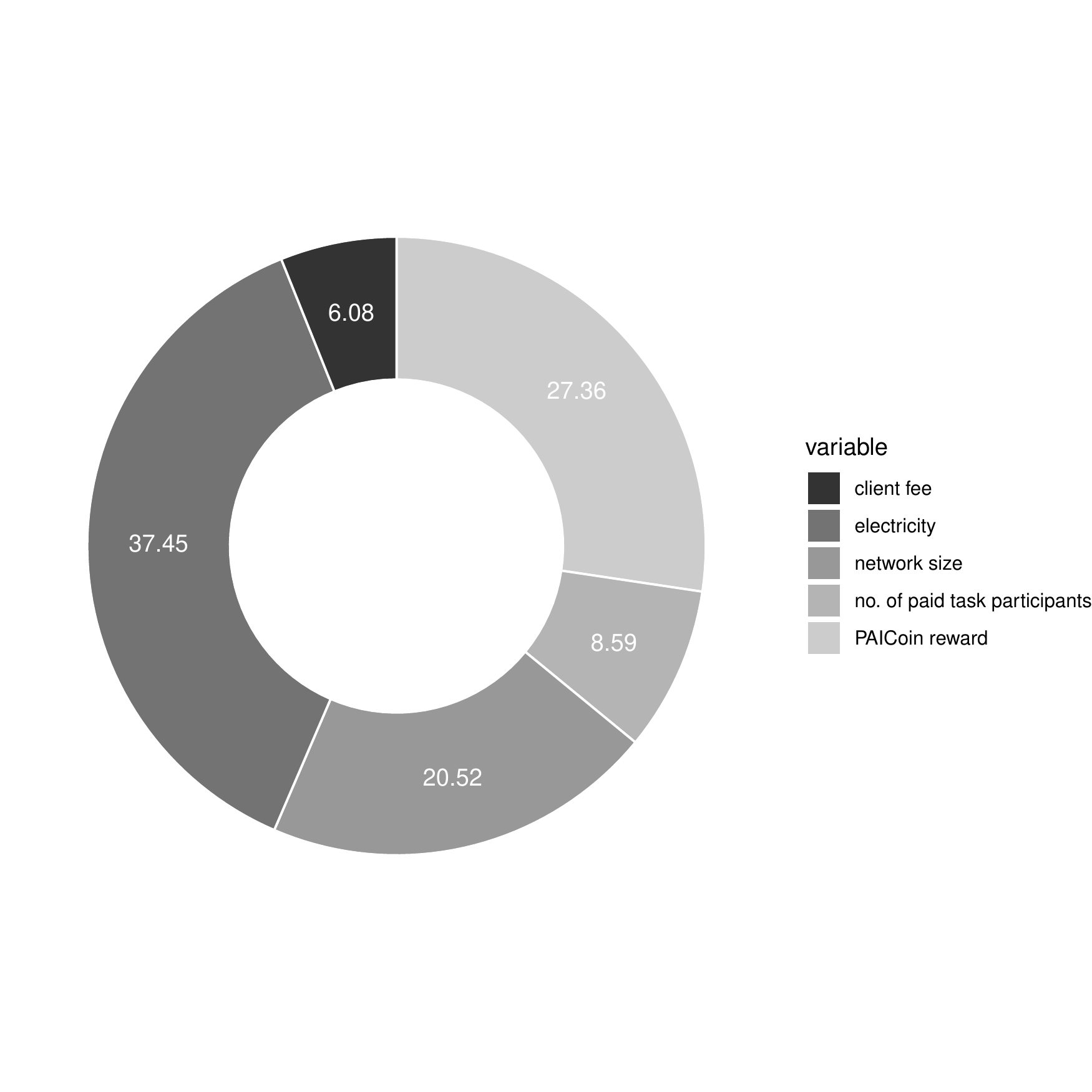}
\caption{\textbf{Importance of variables.} We plotted the most important ROI variables (others are \textless 1\%). Variables were assessed by their absolute values.}
\label{fig:importance}
\end{figure}

\begin{figure}[!t]
\centering
\includegraphics[width=0.48\textwidth]{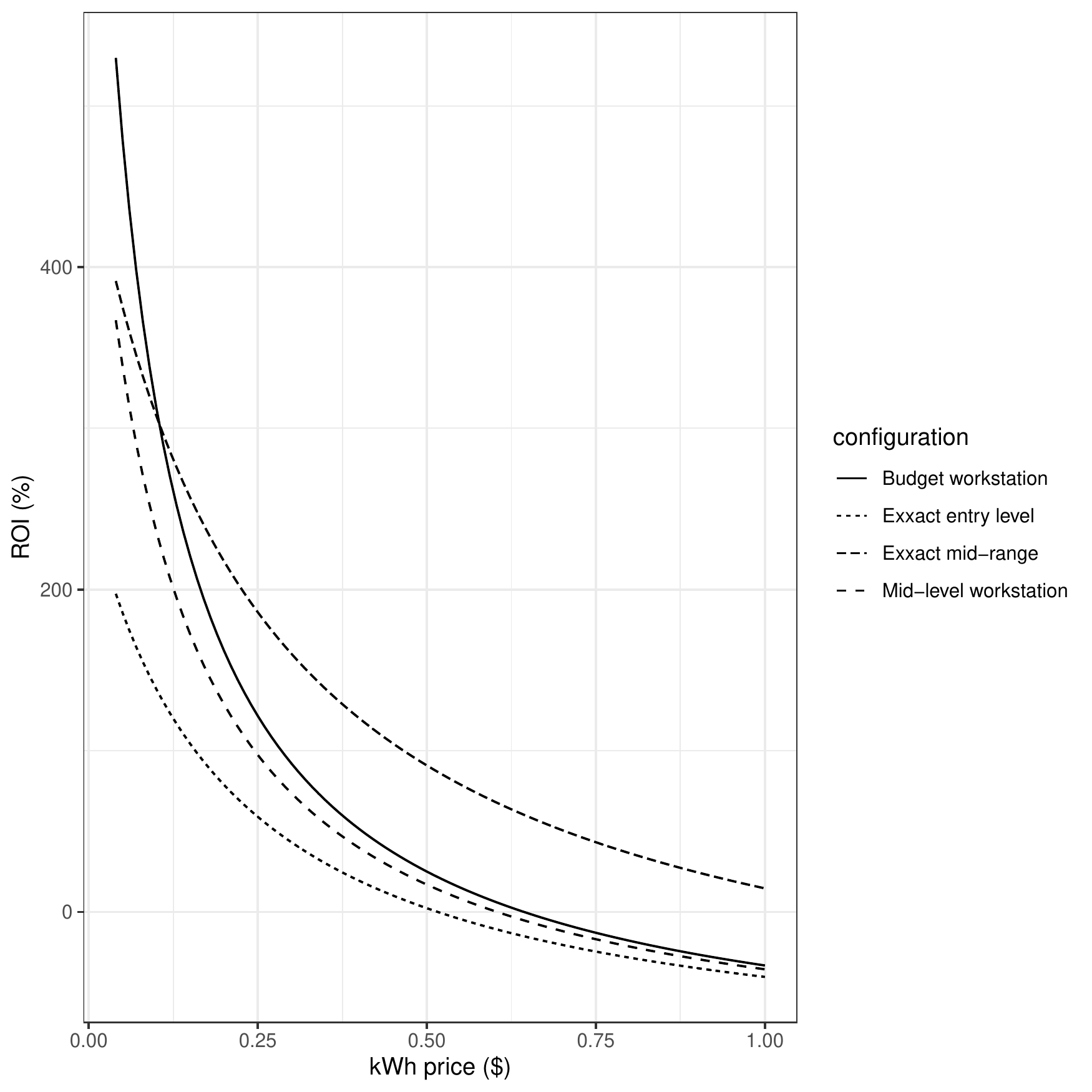}
\caption{\textbf{ROI (\%) as a function of electricity price.}}
\label{fig:electricity}
\end{figure}

To quantify the importance of the ROI variables, we need to take into account real-world constraints: the electricity price is in an interval, the block reward is halving and there are static hardware configurations. In the case when $F_h=0.50$ \$, $Q=20$, $W=1155$ \$  (revenue\_pai\_block=1500 and price\_usd\_paicoin=0.77 \$, the mean all-time price of PAICoin), $U=10000$, $C_H=1945$ \$ and 0.12 \$ for 1 kWh, we have obtained the importance scores (in percentages) outlined in Fig. \ref{fig:importance}. We can see that electricity is the variable whose change has the most influence on the profitability. Fig. \ref{fig:electricity} shows a rapid decrease in profits as the price of a kWh increases. The same is true for the classical Bitcoin. The PAICoin reward (composed of the block reward multiplied by the PAICoin price) is the second most important variable, closely followed by the network size. Other variables that matter are the average number of paid task participants (miners, supervisors and evaluators) and the client's fee.

\begin{figure}[!t]
\centering
\includegraphics[width=0.4\textwidth]{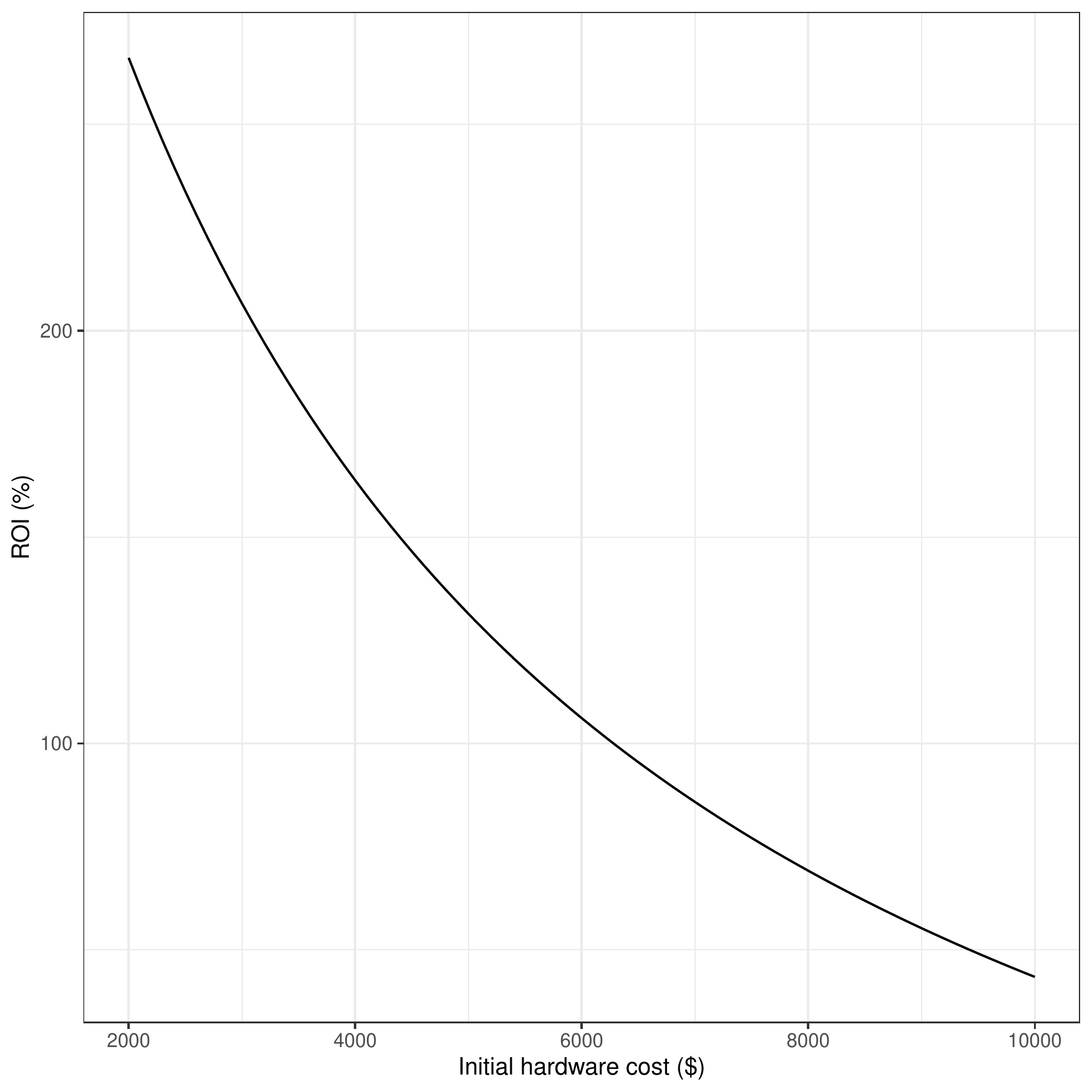}
\caption{\textbf{ROI (\%) as a function of initial hardware cost.}}
\label{fig:hwd_cost}
\end{figure}

However, Fig. \ref{fig:importance} does not capture the impact of the initial hardware investment. From Fig. \ref{fig:hwd_cost}, it might seem that PoUW profitability is higher for miners that invest less in their hardware configurations. To mitigate against this kind of risk, our solution has a matching mechanism between a specific task hardware preferences and the miners' systems capabilities (see Subsection \ref{registration}). More expensive hardware comes with more GPUs ($g>1$), and the miner can participate in several tasks simultaneously (one GPU per task). Additionally, the reward scheme $R$ should allocate a higher compensation for the better hardware.

A miner participates in our network if his/her hourly profit $P_h^{(pai)}$ is positive and greater than the profit obtained by mining Bitcoin using the best mining rig on the market ($P_h^{(btc)}$): $P_h^{(pai)} > P_h^{(btc)}$ and $P_h^{(pai)} > 0$. The hourly profit is calculated as: $P_h^{(pai)} = g\left(\frac{F_h}{Q} + 6 \frac{W}{U}\right) - \left( \frac{C_{H}}{26280} + E \right)$.

Clients use our PoUW system if they pay less than the cheapest cloud option $F_{h_{min}}^{(cloud)}$ (in our case, GCP is $F_{h_{min}}^{(cloud)}=2.36$ \$), attenuated by a variable that describes the distributed system's efficiency, $E_f$, such that: $F_h < F_{h_{min}}^{(cloud)} \cdot E_f$. The advantage (\%) gained by a client using our solution is $100 \left( 1-\frac{F_h}{F_{h_{min}}^{(cloud)} \cdot E_f} \right)$.

With the price of Bitcoin being \$8243.41 (as of October 2019), only the best Bitcoin mining rig on the market (Antminer S17 Pro) achieved a 18\% ROI in our calculations, while the PoUW miner configurations achieved a mean ROI of around 200\% using the previously stated parameter values. Therefore, PAICoin PoUW mining was approx. 10x more profitable than Bitcoin. For clients, we also obtained reduced fees by approx. 30\% vs. using the cheapest cloud ML option.

\subsection{Performance considerations}\label{Performance}

A distributed solution will never be as efficient as running the whole training on a local machine, but there are several optimisations which can improve the performance: miners can receive gradients asynchronously during the whole iteration, the mini-batches can be preloaded, building the message map can be done in parallel. Mining can be performed in a different thread or process. Worker nodes can discard unnecessary data. Only verifiers carry out full verifications, regular peer nodes are not overwhelmed with expensive computations. In our Supplementary Material, we offer more information about performance characteristics and how we can increase the efficiency $E_f$.

To optimise bandwidth, we use the "dead-reckoning" scheme from Strom in \cite{Strom2015}. The technique achieves a reduction of 99.6 \% in data transfers for a model with 14.6 million parameters, as reported in \cite{skymind_2018}.

\subsection{Security}

We assume that the majority of participants is honest. We designed our PoUW protocol to avert various threats from Byzantine actors. We list the strategies to disincentivize and make cheating impossible in the following paragraphs:

\subsubsection*{Client uses pre-trained models}
In this scenario, a client has pre-trained a model with good performance. He/she submits the ML task to the PAI network and then acts as a miner. If the malicious actor is selected to work on the task, he/she could submit his/her pre-trained model without doing ML work, while mining for PAI coins.

Homomorphic encryption would ensure adequate protection, but it is computationally infeasible \cite{rist_2018}. To mitigate this risk, each miner must send weight updates at every iteration and must provide the inputs for lucky nonces. A malicious actor cannot perform cheap work because his/her progress depends on peer updates.

\subsubsection*{Client uses a test set from another distribution}
The client might do this to obtain a trained model without having to pay for it because it would have a bad performance on the test dataset. The training and the test dataset should be identically distributed.

In our protocol, the client submits the whole dataset that is shuffled and split into a training and a test dataset based on the hash of the task definition block (as described in subsection \ref{data prep}). The task definition block is unknown, it doesn't exist before task submission and subsequent mining.

\subsubsection*{Client submits a malformed task definition}
We allow the worker nodes to inspect and report if the task T is malformed. In that case, the client's fee is confiscated and distributed to worker nodes and to evaluators.

\subsubsection*{Gradient poisoning attack}

Gradient poisoning is a type of attack in which a miner tries to skew the learning process by sending huge or fake gradients. Sending the same message multiple times is also a type of poisoning (spam). Blanchard et al. have proven in \cite{Blanchard2017} that only one Byzantine actor could significantly affect the ML process and proposed the \emph{Krum function} for detection. Damaskinos et al. also proposed the \emph{Kardam filter} \cite{Damaskinos2018} for dealing with this threat.

Supervisors watch for this attack and add the malicious worker node to a blacklist which they expose publicly. Fellow nodes will ignore the gradient updates from the bad miners and evaluators will confiscate their stakes. Also, miners will not apply multiple IT\_RES messages corresponding to the same iteration. To turn away miners that do not make progress, validators require that a lucky miner must prove that his/her local model improved over previous iterations.

\subsubsection*{Miner performs only mining}
We constrain the number of nonces to make classical mining insignificant. A miner that would do bogus ML work and focus only on mining would be unable to prove the validity of the produced blocks. Bogus ML work includes: echoing received weight updates, leaving the task before completion or not following the steps in the ML training. It is economically damaging to the miners to engage in such behaviours because they would lose their stake and wouldn't receive any fee from the client anyway.

\subsubsection*{Sybil attacks}
Bad actors can setup several Sybils on the network to collectively generate a bad model. They could also perform cheap work by replicating only one unit of work across all controlled nodes. To avert this attack, worker nodes are not allowed to pick tasks themselves, they only state their preferences (see Subsection \ref{registration}). By doing so, they also cannot pick easy tasks.

\subsubsection*{Byzantine leader}
If the leader of the supervisors delays publishing the \emph{MESSAGE\_HISTORY} transactions, then another leader is immediately elected. If the leader publishes invalid MESSAGE\_HISTORY transactions, then he/she is added to the blacklist and replaced.

\subsubsection*{DOS attacks}
Worker nodes may suspect that a DOS attack takes place when they do not receive enough peer updates or when the training process is very slow (or stalled). They can pause the training process and resume it when the attack is over.

The procedure to defend against DOS attacks is: several worker nodes issue a \textit{CONSIDER\_PAUSE} transaction with the reason \textit{DOS\_ATTACK}. When a majority is considering to pause within a predefined time-frame, then honest workers emit a \textit{PAUSE} transaction and everybody pauses. In pause mode, every concerned node will send \textit{HEARTBEAT} off-chain messages to the former nodes in the ML task group. When a new majority of active nodes is formed, they can publish \textit{CONSIDER\_RESUME} messages and finally, \textit{RESUME} transactions to continue the training process.

A verifier can reject a block mined during a DOS attack if there are enough elements to suspect that it was mined by an attacker.

\subsubsection*{Blockchain spam}
A malicious miner could flood the blockchain with bogus blocks and determine honest verifiers to spend a considerable effort to validate them. This is also a DoS attack because it is hard to validate these blocks very fast. We adopt the following countermeasures:
\begin{itemize}[leftmargin=*]
	\item We prioritise the less expensive verification operations to be run first.
	\item A miner's stake gets confiscated and the miner is blacklisted if he/she submits invalid blocks.
	\item We limit the number of blocks a miner can publish during a predefined time interval.
\end{itemize}

\subsubsection*{Long-range attacks}
In a \emph{long range attack}, an attacker forks a large number of blocks or the entire blockchain starting with the genesis block. If the attacker has a high computational power, he/she can even outpace the main chain and publish his/her alternative chain.

Our underlying blockchain is a hybrid PoS/PoW blockchain protected by the longest chain rule. New blocks are created by spending a considerable amount of energy on useful work. We also require that at every 1024 blocks a checkpoint is created: everything before the checkpoint is truly immutable (i.e. no change can be done later to parts of the blockchain deeper than 1024 blocks). To establish a checkpoint, a set of 12 validators are randomly chosen to vote. At least 9 out of 12 votes are needed to establish a checkpoint.

\section{Conclusion} \label{Conclusion}
We presented a novel proof of useful work concept using a distributed and decentralised machine learning system on blockchain. Our proposal can be easily extended to other AI algorithms.

We briefly reviewed the related work and outlined the unique characteristics of our solution. We conceived a different blockchain framework in which miners get compensation for doing useful work and we elaborated mechanisms to deter and punish bad actors. We presented the roles, the environment and the consensus protocol that combines machine learning with blockchain mining to create and reward useful work.

In our system, a client can train a ML model using a distributed network of worker nodes. After they perform a small unit of pre-assigned work, miners can mine new blocks with special \emph{nonces}. At each iteration, nonces are obtained with a formula that takes into consideration inputs and by-products of the ML training. If a miner finds a lucky nonce, he/she must prove that he/she executed honestly the iteration so that his block will be accepted by the rest of the network. Verification means re-running the lucky iteration. Because miners are using data parallelism to train their models, they need to exchange information quickly using off-chain messages. Most of these messages carry data about updates that a miner performed to his/her local model. These updates are replicated across the task group by fellow worker nodes. A message history is recorded and will serve later in the verification. Compared to other blockchains, a node always receives compensation from the client; solving the blockchain puzzle is a bonus. Although we constrain the nonces to several values, the target difficulty in the network is very low in order to mine a block every 10 minutes as in the Bitcoin protocol. We shift the mining process towards ML training, while the actual hashing is insignificant.

We also implemented a proof of concept for PoUW. We showed that our PoUW solution is more cost-friendly to a client than regular cloud ML training, but also more profitable to miners compared to Bitcoin mining. Our approach also shows that ML models can be trained collectively with good performance using commodity hardware owned by individuals. We believe that such a system would democratise artificial intelligence using the security of the blockchain technologies. 

In this paper, we described the particular case of training a deep neural network (DNN), but the principles can easily generalise to most AI iteration-based algorithms. Future work includes adding multi-party secure computation to the protocol and a production-ready implementation of the system described in this paper.

\section*{Acknowledgments}

The authors would like to thank Muhammad Naveed, Assistant Professor of Computer Science at the University of Southern California for his extraordinary assistance and for reviewing this work.

\nocite{*}

\bibliographystyle{compj}

\bibliography{ms}

\appendix
\onecolumn
\section{Anatomy of transactions}\label{appendix_transactions}

PAI nodes recognise transaction types by reading their last TXOUT’s. They contain descriptions in their OP\_RETURN scripts. Here is an example of a possible implementation. Details are given in Tables \ref{tab:anatomy1}, \ref{tab:anatomy2} and \ref{tab:anatomy3}.

\begin{table}
\caption{BUY\_TICKETS}
\label{tab:anatomy1}
\centering
\begin{tabular}{{|p{1.8cm}|p{10cm}|}}
\toprule
\rowcolor[HTML]{C0C0C0} 
	transaction & description \\ \midrule
	txin[0..n] &  references to existing UTXOs   \\ \midrule
	txout[0] &  contains a value that is a multiple of the current price of tickets
           of the according type (miner, supervisor etc.); script field is empty!   \\ \midrule
	txout[1] &  returns change to the staker as P2PKH;
           this is the address to which mining remuneration will be paid   \\ \midrule
	txout[2] &  an OP\_RETURN that declares ticket type and preferences: "TICKET-TYPE:preferences"   \\ \bottomrule
\end{tabular}
\end{table}

\begin{table}
\caption{PAY\_FOR\_TASK}
\label{tab:anatomy2}
\centering
\begin{tabular}{{|p{1.8cm}|p{10cm}|}}
\toprule
\rowcolor[HTML]{C0C0C0} 
	transaction & description \\ \midrule
	txin[0..n] &  these reference existing UTXOs   \\ \midrule
	txout[0] &  declaring only the amount staked; script field is empty!   \\ \midrule
	txout[1] &  returns change to the staker   \\ \midrule
	txout[2] &  an OP\_RETURN that describes the task   \\ \bottomrule
\end{tabular}
\end{table}

\begin{table}
\caption{CHARGE\_FOR\_TASK}
\label{tab:anatomy3}
\centering
\begin{tabular}{{|p{1.8cm}|p{10cm}|}}
\toprule
\rowcolor[HTML]{C0C0C0} 
	transaction & description \\ \midrule
	txin[0..n] &  these reference task-related stakes (client's stake as well as captured stakes 
             of participants who transgressed); script fields are empty!   \\ \midrule
	txout[0..n] &  these are classic P2PKH outputs that pay participants; note that only 
              one CHARGE\_FOR\_TASK transaction is spendable   \\ \midrule
	txout[n+1] &  an OP\_RETURN script that publishes the best model for the
             client to download: "RESULT:taskHash:url"   \\ \midrule
\end{tabular}
\end{table}

The specification of the Bitcoin OP\_RETURN outputs states that the first opcode is OP\_RETURN and it is followed by a push opcode. We introduced a richer format called \emph{structured data outputs}, which are recognized as having OP\_RETURN fields followed by special OP\_STRUCT opcodes, which in turn are followed by an array of data items. The specification for the structured data format also states that the first data item must specify the version, followed by transaction specifics. For example, for PAY\_FOR\_TASK, the next fields contain details about the dataset, validation, optimiser etc.

\section{Shortening task wait time}\label{shortening}

We aim to shorten the waiting time of starting a task. We require that the timestamp of the task is earlier than the block containing it and the task must eventually appear in all blockchain forks.

A node keeps a local list of tasks along with their detection time (which differs across nodes by seconds). To add new elements to it, each node searches the chain starting from task's timestamp, including orphaned forks in the traversal. Any mined block containing the task submission is a task definition block, but the nodes should recognise only one of them. Therefore, the oldest block is chosen, regardless of the branch. In case there is more than one block with an identical timestamp containing the same task submission, the one with the lowest hash is picked.

Selected workers start joining by sending JOIN\_TASK transactions, in which they include the hash of their perceived task definition block and the task hash. A majority with a specific task definition hash will emerge and only those selected by it will start the training. Those who are not selected are ignored if they try to participate.

When training is completed, the part of the ledger around task submission moment will have been stabilised. Nodes have to find the task stake block to reference the PAY\_FOR\_TASK stake. They search the chain starting from the task's timestamp, traversing only the active branch and disregarding stalled forks. The first block containing the task submission is the task stake block.

Using this procedure the network can safely start working on tasks long before the stakes are confirmed.

\section{A Distributed Key Generation (DKG) scheme} \label{dkg}

In our PoUW system, the private key shares are constructed using a \emph{distributed key generation} (DKG) protocol during task initialisation, which is a modified version of the Joint-Feldman protocol (\cite{Pedersen:1991:TCW:1754868.1754929}). It contains the following steps:
\begin{itemize}[leftmargin=*]
	\item Each supervisor creates an $t-1$-degree polynomial $\mathcal{S}_i(x)= s_{i,0} + s_{i,1} x+ ...+s_{i,t-1}x^{t-1}$, where all the coefficients except $s_{i,0}$ are randomly generated private keys. $s_{i,0} = s_{k_i}$ is the private BLS key of the participant.
	\item Based on their $\mathcal{S}_i(x)$, all supervisors calculate and broadcast their own $\mathcal{P}_i(x)=p_{i,0} + p_{i,1} x + ... + p_{i,t-1}x^{t-1}$ which is another polynomial of the same degree, that holds the corresponding public keys, such that $p_{i,j} = g_i \times s_{i,j}, j \in \{0,..t-1\}$.
	\item Every party evaluates $\mathcal{S}(x)$ for all participants and for itself, by replacing $x$ with the corresponding index of each supervisor. For example, in case we are dealing with a 3-of-5 scheme, supervisor 2 will calculate the following:
	\begin{gather*}
    \mathcal{S}_2(1)=s_{2,0} + s_{2,1} \cdot 1 + s_{2,2} \cdot 1 \\ 
    \mathbf{\mathcal{S}_2(2)=s_{2,0} + s_{2,1} \cdot 2 + s_{2,2} \cdot 2^2} \\
    \mathcal{S}_2(3)=s_{2,0} + s_{2,1} \cdot 3 + s_{2,2} \cdot 3^2 \\
    \mathcal{S}_2(4)=s_{2,0} + s_{2,1} \cdot 4 + s_{2,2} \cdot 4^2 \\
    \mathcal{S}_2(5)=s_{2,0} + s_{2,1} \cdot 5 + s_{2,2} \cdot 5^2
\end{gather*}

	\item Every supervisor will encrypt and send every $\mathcal{S}_i(x)$ to the corresponding parties. E.g. supervisor 2 will use supervisor 1's public key to encrypt $\mathcal{S}_2(1)$ and send it to supervisor 1. In case one party does not receive all the private shares during a pre-set time window, he/she will complain against the senders by sending \textit{DKG\_COMPLAINT} transactions to the blockchain containing the indexes of the senders.
	\item Every private share is decrypted upon arrival and verified using $\mathcal{P}(x)$ by replacing $x$ with his/her own identifier/index. If the result does not match the public key derived from the received private share, the node will fill a complaint against the sender by publishing a \textit{DKG\_COMPLAINT} transaction on the blockchain, that contains the identity of the culprit and the received private share.
	\item To obtain the global public key ($\mathbf{\mathcal{P}_k}$), all supervisors will aggregate the free terms of all polynomials $\mathcal{P}(x)$: $\mathbf{\mathcal{P}_k} = \sum_{i=1}^{n} p_{i,0} = p_{1,0} + p_{2,0} + ... + p_{n,0}$, which are publicly known.
	\item The global private key $\mathbf{\mathcal{S}_k} = \sum_{i=1}^{n} s_{i,0} = s_{1,0} + s_{2,0} + ... + s_{n,0}$ is unknown to any party.
	\item At the end of the protocol, all supervisors must post \textit{DKG\_SUCCESSFUL} transactions containing the locally calculated \emph{t-of-n} public key $\mathbf{\mathcal{P}_k}$. When $n$ such transactions with the same public key are observed during a predefined time window $\Delta t$, then the DKG protocol is considered successful and the parties can proceed to the next phase.
\end{itemize}

A DKG protocol runs in the initialisation phase of a ML task (after the key exchange), but also whenever the supervisory committee changes. Supervisors that produce faults during DKG are banned from the network and their stakes are confiscated.

It is not easy to detect if a node received a wrong share or the node received a correct share but pretends that he/she didn't in order to exclude another node. If 2/3 of the nodes complain against one node, then that supervisor is automatically disqualified; however, if only one node complains against another, then the other nodes will vote based on the current evidence which of the nodes will be excluded: the sender or the receiver. The DKG is restarted with the existing non-faulty members if their reduced number still satisfies the size requirements or with new members replacing faulty ones if there would be less worker nodes than required. The selection procedure for additional worker nodes is the same as the one provided in the task registration phase. Miners will include JOIN\_TASK transactions for the extra nodes in a second participation block that should reference the first participation block.

\section{Signing \emph{t-of-n} transactions}\label{t_n}
Each supervisor has several secret key shares that are used to sign \emph{t-of-n} transactions. For a transaction $tx$, each supervisor will follow these steps:

\begin{enumerate}
	\item Compute $H(tx)$, which is the hash of the transaction to the BLS curve.
	\item Publishes ${Sig}_i(tx)=\sum_{j=1}^n \mathcal{S}_j(i) \times H(tx)$, where $i$ is the index of the current node and $j$ indexes the private shares. Please note that each node $i$ calculates the aggregation of its private key shares as $\sum_{j=1}^n \mathcal{S}_j(i)$ and uses it to sign the transactions.
	\item The epoch leader collects at least $t$ signature shares.
\end{enumerate}

For a given transaction $tx$, as soon as any $t$ signature shares are collected, due to the BLS threshold signature properties, the leader can reconstitute the global signature on the transaction ($Sig(tx)$) by performing Lagrange interpolation, as if the global private had been used to sign the transaction ($Sig(tx) = (s_{1,0} + s_{2,0} + ... + s_{n,0}) \times H(tx)$). The global signature validates against the free coefficient of the global public key $\mathcal{P}(x)$.

\section{Crash-recovery model} \label{Crash}
We use a crash-recovery model to detect when and which nodes go offline. It is inspired from \cite{Backes03reliablebroadcast}, a framework in which nodes may crash and recover repeatedly, while some may go offline permanently. In real-life, although there are potential network problems or software/hardware glitches, nodes eventually go back online and continue the ML training.

\subsection{Offline detection}
If a worker node suspects that another worker node (miner or supervisor) becomes too slow or does not send the expected messages in a reasonable amount of time $t_r$, then he/she launches a test to detect if the node is offline (crashed). The probing algorithm (Algorithm \ref{alg:faildetect}) is inspired from an algorithm called SWIM (\cite{Das:2002:SSW:647883.738420}). We modified the algorithm to become Byzantine Fault Tolerant (BFT). As in PBFT (\cite{Castro:1999:PBF:296806.296824}), we require that more than 1/3 of the nodes (the maximum accepted number of possible faulty nodes in a BFT system) should declare that a particular node is online so that the entire task group considers that the node is online. We assume a weak synchrony as in \cite{Castro:1999:PBF:296806.296824}, i.e. network faults are eventually repaired and most nodes are coming back from the offline mode fairly quickly.

Any worker node $w_i$ keeps a local list of known active worker nodes $W$. As outlined in Algorithm \ref{alg:faildetect}, a supervisor $s_i$ pings another suspected worker node $w_j$. If no response is received in a time interval $\Delta t$ then $s_i$ will randomly select a subset of $k$ worker nodes ($W_k \subsetneq W$) nodes and ask them in parallel to also ping the node in question. To avoid any bias, the $k$ chosen nodes are determined by a random number generator seeded with the hash of the last participation block. The number of "alive" responses are counted. If more than 1/3 of the enquired nodes declare that the node is alive then all nodes must keep it in their local list. Each response is signed by the sender. Otherwise, the suspecting worker node will publicly contact the leader and ask him/her to decide whether the node should be replaced (ReportOfflineNode procedure). To do so, the suspecting node will publish a transaction (\emph{CHECK\_NODE}) containing the responses from peers and the reason for investigation (e.g. "offline"). The leader runs an election in which supervisors must vote on whether the node should be declared offline or online. Each supervisor will directly ping the suspected offline node and if no response is received in a specified time interval, then the supervisor will vote it as offline. The leader will publish a \emph{NODE\_STATUS} \emph{t-of-n} transaction with all the votes. If the final status is offline (2/3 or more of the supervisors voted "offline"), then the leader will publish a \emph{RECRUIT\_WORKER\_NODE} transaction containing the ID of the replaced worker node and the reason for replacement. The working nodes will remove the offline node from their lists.

Algorithm \ref{alg:investigate} is similar to the Algorithm \ref{alg:faildetect}, but it is run only by supervisors.

A node that is going in offline-online mode too often must be removed from the task working group using the \emph{CHECK\_NODE} and \emph{NODE\_STATUS} transactions, with "offline-online" as the reason. Malicious nodes are reported and verified in the same way using different reasons (e.g. "denial-of-service attack", "gradient poisoning" -- sending wrong updates to derail the ML training etc.).

\begin{algorithm}
\caption{Algorithm used by a worker node $w_i$ to detect if a suspected node $w_j$ is offline.}\label{alg:faildetect}
\begin{algorithmic}[1]
\Procedure{DetectOfflineNode(...)}{}
\State $r \xleftarrow{\Delta t} ping(w_j)$ \Comment Ping $w_j$ and wait for a time $\Delta t$ for a response
\If{$r == online$}
    \State $W \gets W \cup w_j$ \Comment Keep or add it to the list
\Else \Comment No response
    \State $W_k \gets rnd(W \setminus \{w_i, w_j\})$ \Comment Pick $k$ random worker nodes (a subset of $W$).
	\State $c \gets 0$ \Comment Online counter.
	\State $R \gets \{r\}$ \Comment Responses.
    \ParFor{$w_k\in W_k$} \Comment In parallel.
        \State $R_k \xleftarrow{\Delta t} ping\_req(w_k)$ \Comment Send ping requests to each chosen node.
        \If{$status(R_k) == online$}
            \State $c \gets c + 1$
        \EndIf
    \EndParFor
    \If{$c > |W|/3 $} \Comment If more than 1/3 report the node as online
        \State $W \gets W \cup w_j$ \Comment The node is alive.
    \Else
    	\State $s \gets ReportOfflineNode(w_j, W_k, R)$ \Comment Report the $k$ chosen nodes and the responses.
    	\If{$s$ == offline}
             \State $W \gets W \setminus w_j$ \Comment Remove the node from the active list.
        \EndIf
    \EndIf
\EndIf
\EndProcedure
\end{algorithmic}
\end{algorithm}

\begin{algorithm}
\caption{Algorithm run by the leader to determine if he/she should declare a worker node as offline.}\label{alg:investigate}
\begin{algorithmic}[1]
\Procedure{InvestigateNodeStatus(...)}{}
\State $c_x \gets 0$ \Comment Offline counter.
\State $r \xleftarrow{\Delta t} ping(w_j)$ \Comment Ping $w_j$ and wait for a time $\Delta t$ for a response
\If{$r \neq online$}
    \State $c_x \gets 1$ \Comment Offline counter.
\EndIf
\State $R \gets \{r\}$ \Comment Responses.
\ParFor{$s_k\in S_k \setminus \{s_i\}$} \Comment In parallel.
        \State $R_k \xleftarrow{\Delta t} ping\_req(s_k)$ \Comment Send ping requests to each supervisor.
        \If{$status(R_k) == offline$}
            \State $c_x \gets c_x + 1$
        \EndIf
    \EndParFor
\If{$c_x \geq 2/3 |S|$} \Comment If more than 2/3 report the node as offline
        \State $Tr(RECRUIT\_WORKER\_NODE, R)$ \Comment Issue transaction to recruit a new worker node
         \Else 
 		\State $Tr(NODE\_STATUS\_ONLINE, R)$
 \EndIf
\EndProcedure
\end{algorithmic}
\end{algorithm}

\subsection{Offline supervisors} \label{Offline supervisor}
A supervisor could go offline because of a faulty network connection. If he/she loses more than 10\% of the training iterations, then he/she loses his/her stake and cannot rejoin the task.

If a supervisor comes back online and he/she didn't lose over 10\% of the iterations, then he/she can synchronise with the other supervisors. The supervisor will read from the public streams/databases of the other supervisors and will get up-to-date.

If a supervisor is missing for periods of more than 10\% of the training, then the other supervisors will initiate a recruitment process to replace the absent supervisor. The leader will post a \emph{RECRUIT\_WORKER\_NODE} transaction which will specify that a new supervisor is needed.

\subsection{Offline miners}
If a miner does not participate in more than 5\% of all training operations, he/she will lose his/her stake. Otherwise, he/she can rejoin the task. No model synchronisation is needed, yet the miner has to catch up with the latest weight updates.

Another miner is called in by the supervisors when the total number of miners drops to less than 80\% of the original size. The leader will publish a special transaction called \emph{RECRUIT\_WORKER\_NODE}. The stake of the lost miner will be transferred to the task's stake and redistributed by the evaluators at the end of the training.

\subsection{Recovery}
A supervisor or miner that comes back online but did not lose enough iterations in order to be replaces will synchronise his/her internal database with the rest of the group. No voting to re-accept the old member is required.

\section{Performance of the ML distributed system}
\begin{figure}[ht]
    \centering
    \includegraphics[width=0.85\textwidth]{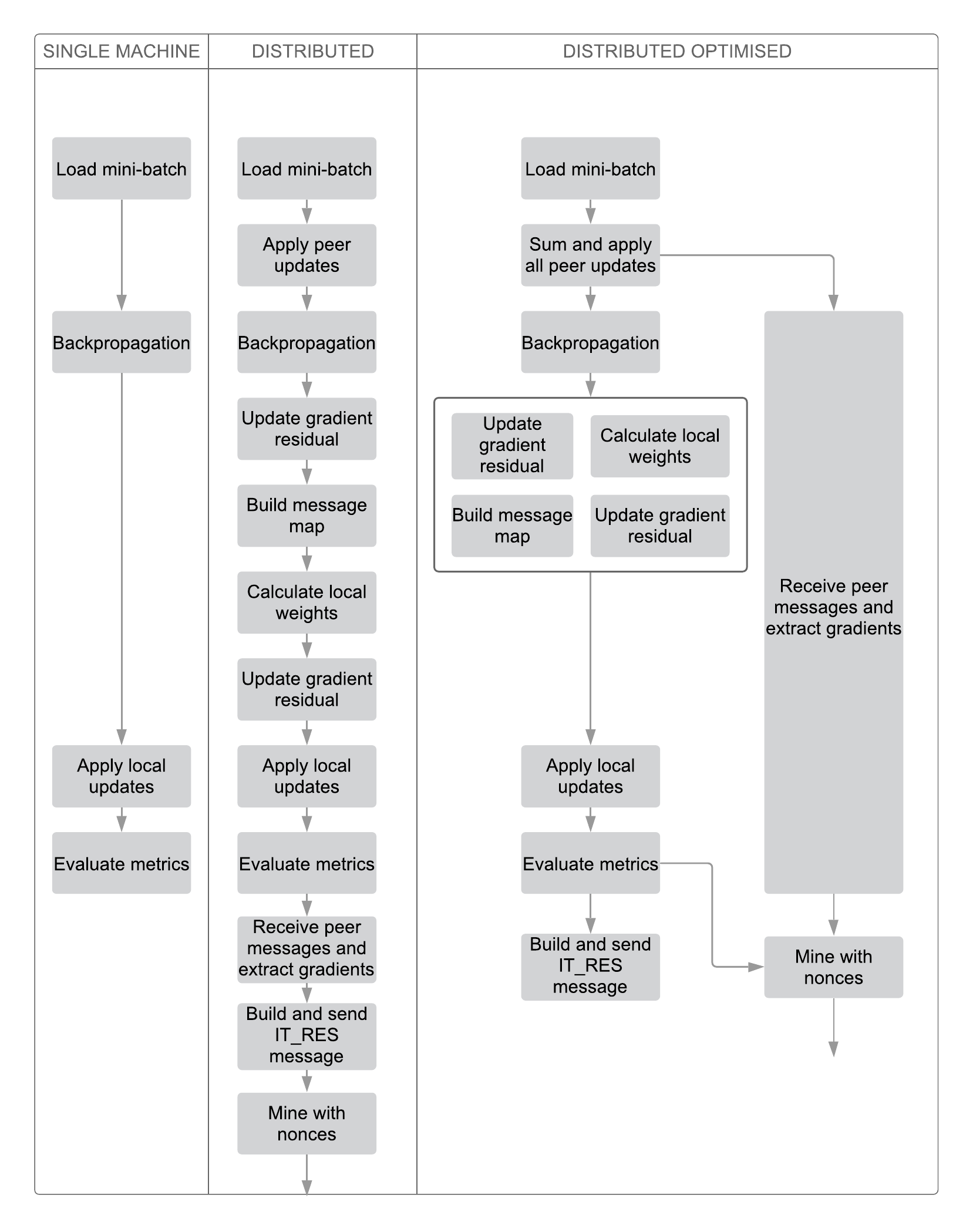}
    \caption{\textbf{Performance optimisations.}}
    \label{fig:efficiency}
\end{figure}

The efficiency of a ML distributed system is less than the one of running the training on a single machine. We consider $E_f=1$ when the training is run entirely on a single machine. In case of a distributed system, $E_f$ is much lower, but can be increased by parallelising different operations, as shown in \ref{fig:efficiency}.

In the left pane of the figure, we show the operations that would take place on a single machine. In the middle, we have an unoptimised scenario in which all operations are executed serially. In the right pane, we improved the execution: peer messages are asynchronously received during the whole iteration, the peer updates are summed and applied once, while the message map is done concomitantly with other related operations.

\begin{figure}[ht]
\centering
  \begin{subfigure}[c]{.67\textwidth}
    \includegraphics[width=.8\linewidth]{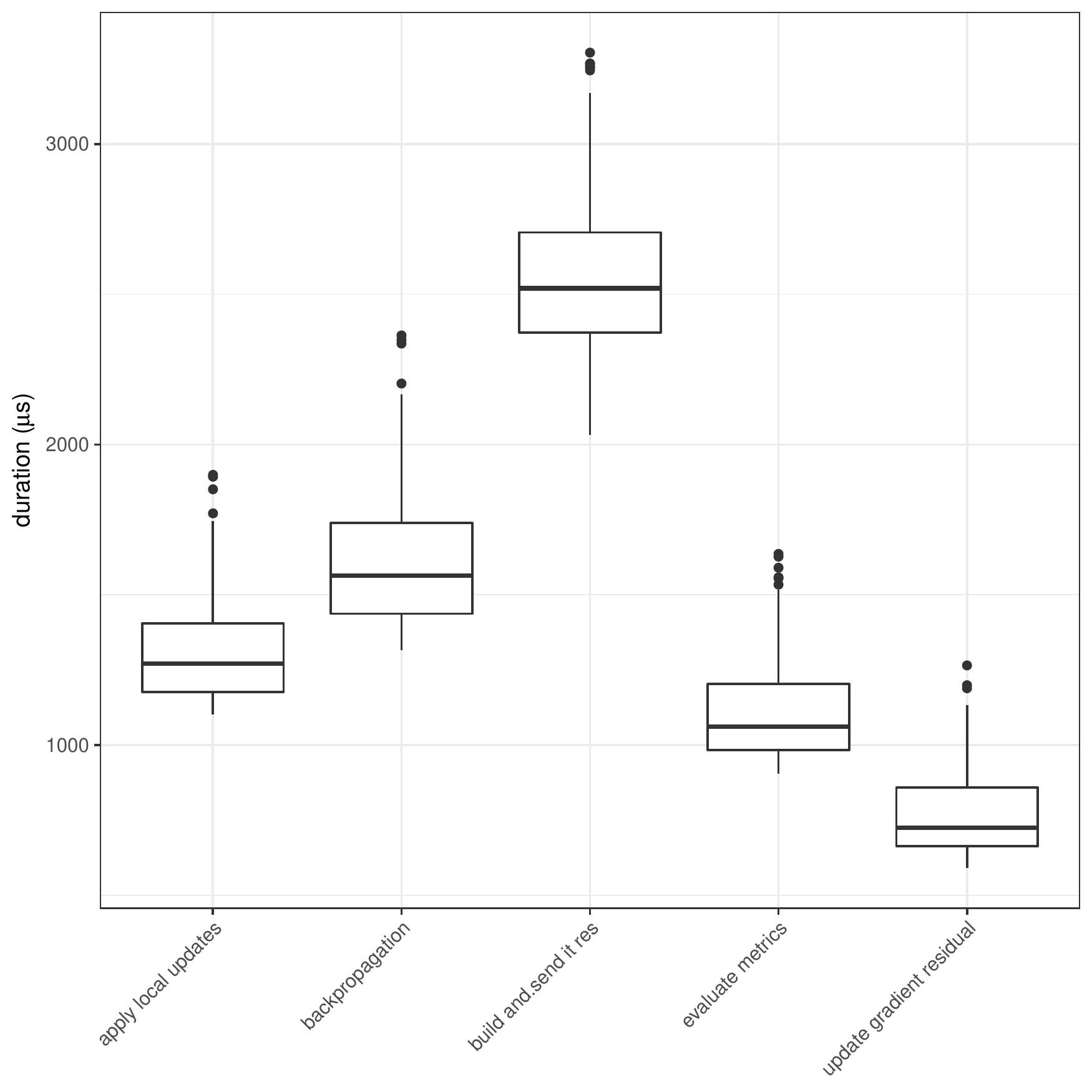}
  \end{subfigure}
  \begin{subfigure}[c]{.67\textwidth}
    \includegraphics[width=.8\linewidth]{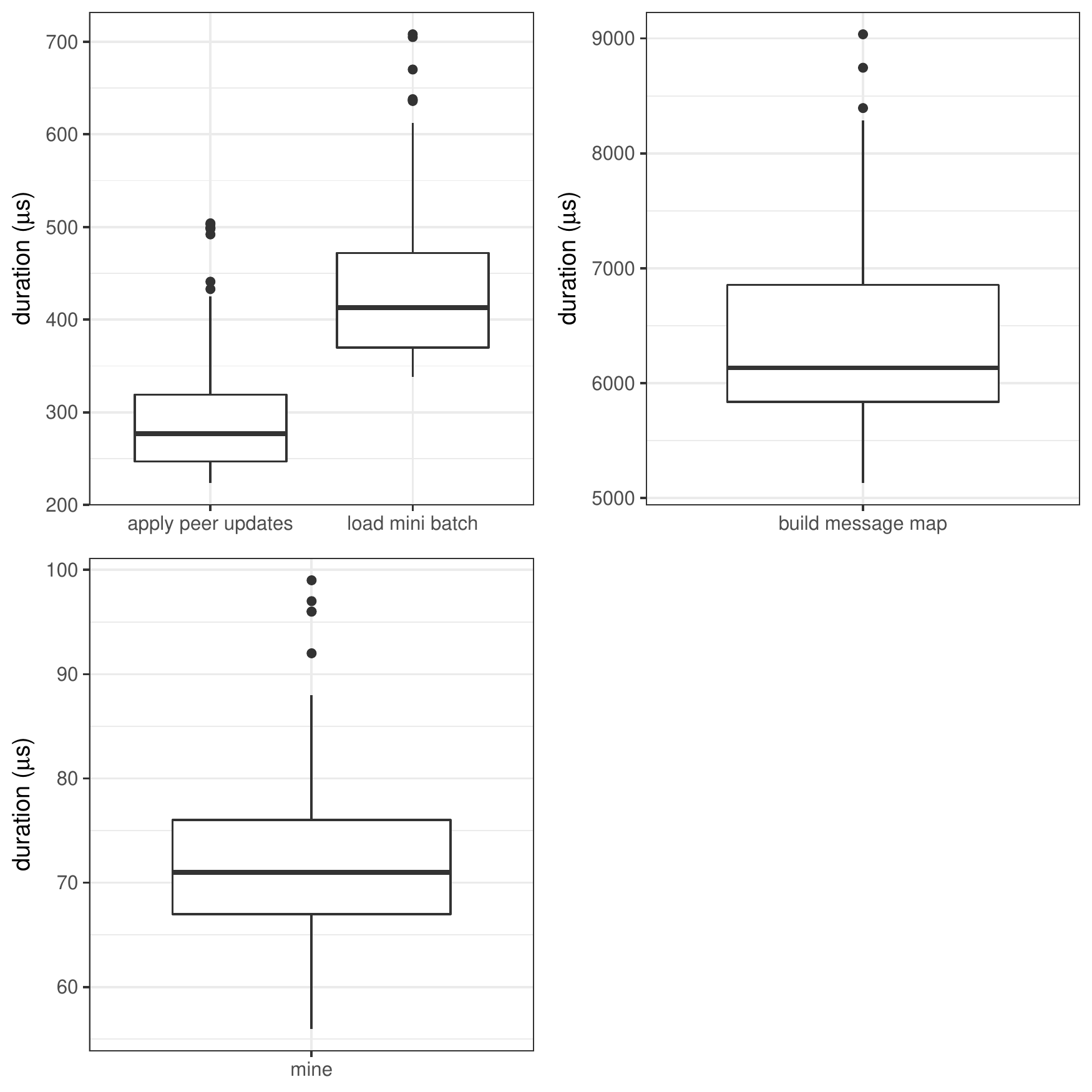}
  \end{subfigure}
  \caption{\textbf{Time measurements for different steps on CPU.}}
  \label{fig:cpu_steps}
\end{figure}

\begin{figure}[ht]
    \centering
    \includegraphics[width=0.9\textwidth]{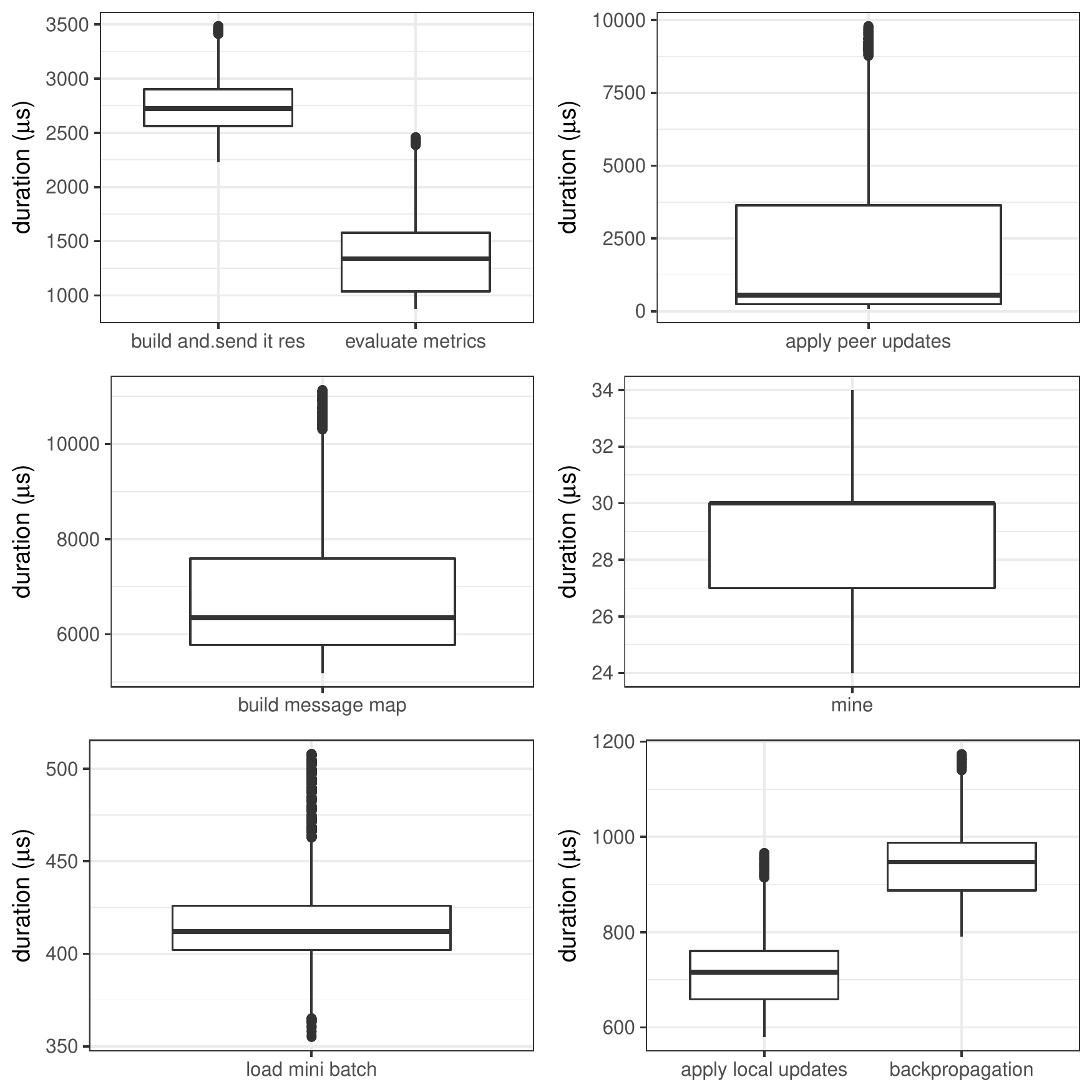}
    \caption{\textbf{Time measurements for different steps on GPU.}}
    \label{fig:gpu_steps}
\end{figure}

We measured the execution times for various steps in the main algorithm on a machine that used only the CPU and on another machine equipped with a GPU. The first machine was a MacBook Pro 2017 with 2,8 GHz Quad-Core Intel Core i7 CPU and 16 GB of RAM. The GPU machine was a NC6 Microsft Azure machine (E5-2690v3 Intel Xeon CPU with 56 GB RAM and 1 x K80 GPU). We provided the measurements in \ref{fig:cpu_steps} and \ref{fig:gpu_steps}, respectively. As expected, most of the steps that involve ML operations (e.g. backpropagation, model updates) are executed faster on the GPU machine.

\end{document}